\documentstyle[emulateapj]{article}

\def\etal{{\rm et~al.\ }}
\def\hmpc{\;h^{-1}{\rm Mpc}}
\def\invhmpc{\;h\;{\rm Mpc}^{-1}}

\def\kms{{\rm \;km\;s^{-1}}}
\def\invkms{({\rm km\;s}^{-1})^{-1}}
\def\kmsmpc{\kms\;{\rm Mpc}^{-1}}
\def\lya{Ly$\alpha$}
\def\lyb{Ly$\beta$}

\def\nh1{n_{\rm HI}}
\def\taueff{\overline{\tau}_{\rm eff}}
\def\deltaf{\Delta^{2}_{\rm F}(k)}

\def\K{{\rm K}}

\def\p1dk{P_{\rm 1D}(k)}

\begin{document}

\submitted{Submitted to ApJ, September 29, 1998}
 
\title{The Power Spectrum of Mass Fluctuations 
Measured from the Lyman-alpha Forest at Redshift z=2.5}

\author{Rupert A.C. Croft$^{1,2}$,
David H. Weinberg$^{1}$,
Max Pettini$^{3}$,  Lars Hernquist$^{2,4}$
and Neal Katz$^{5}$} 

\footnotetext[1]{Department of Astronomy, The Ohio State University,
Columbus, OH 43210; racc,dhw@astronomy.ohio-state.edu}
\footnotetext[2]{Harvard-Smithsonian Center for Astrophysics, 
Cambridge, MA 02138; lars@cfa.harvard.edu}
\footnotetext[3]{Royal Greenwich Observatory, Madingley Road, Cambridge,
CB3 OEZ, UK; pettini@ast.cam.ac.uk}
\footnotetext[4]{Lick Observatory, University of California, Santa Cruz,
CA 95064}
\footnotetext[5]{Department of Physics and Astronomy, 
University of Massachusetts, Amherst, MA, 01003;
nsk@kaka.phast.umass.edu}
 
\begin{abstract}
We measure the linear power spectrum of mass density fluctuations at redshift 
$z=2.5$ from the \lya\ forest absorption in a sample of 19 QSO spectra, using 
the method introduced by Croft \etal (1998).  The $P(k)$ measurement covers
the range $2\pi/k \sim 450-2350\;\kms$ ($2-12$ comoving$\hmpc$ for $\Omega=1$),
limited on the upper end by uncertainty in fitting the unabsorbed QSO continuum
and on the lower end by finite spectral resolution ($0.8-2.3$\AA\ FWHM)
and by non-linear dynamical effects.  We examine a number of possible sources
of systematic error and find none that are significant on these scales. In
particular, we show that spatial variations in the UV background caused by
the discreteness of the source population should have negligible effect on
our $P(k)$ measurement.  We estimate statistical errors by dividing the data
set into ten subsamples.  The statistical uncertainty in the rms mass 
fluctuation amplitude, $\sigma\propto\sqrt{P(k)}$, is $\sim 20\%$, and is
dominated by the finite number of spectra in the sample.  We obtain consistent
$P(k)$ measurements (with larger statistical uncertainties) from the high and
low redshift halves of the data set, and from an entirely independent sample of
nine QSO spectra with mean redshift $z=2.1$.

A power law fit to our results yields a logarithmic slope $n=-2.25 \pm 0.18$
and an amplitude $\Delta^2_\rho(k_p)=0.57^{+0.26}_{-0.18}$, where 
$\Delta^2_\rho$ is the contribution to the density variance from a unit 
interval of $\ln k$ and $k_p=0.008\;\invkms$.
Direct comparison of our mass $P(k)$ to the measured clustering of Lyman Break
Galaxies shows that they are a highly biased population, with a bias factor
$b \sim 2-5$.  The slope of the linear $P(k)$, never previously measured
on these scales, is close to that predicted by models based on inflation
and Cold Dark Matter (CDM).  The $P(k)$ amplitude is consistent with some 
scale-invariant, COBE-normalized CDM models (e.g., an open model with 
$\Omega_0=0.4$) and inconsistent with others (e.g., $\Omega=1$).  Even with 
limited dynamic range and substantial statistical uncertainty,
a measurement of $P(k)$ that has no unknown ``bias factors'' offers many
opportunities for testing theories of structure formation and constraining
cosmological parameters.
\end{abstract}
 

\section{Introduction}
Much of modern cosmology is based on the hypothesis that structure in
our Universe arose from the action of gravity on small initial density
perturbations. The power
spectrum of these initial fluctuations, $P(k)$, is a fundamental prediction of
different cosmological theories. Indeed, in the most common models, the initial
Fourier amplitudes of the density are distributed in a Gaussian random fashion,
and $P(k)$ specifies the statistical properties of the initial density 
distribution entirely. A  determination of $P(k)$ would therefore offer a
direct way to test these theories, and to constrain any free parameters 
they might have. Also, and perhaps just as importantly, an unambiguous
measurement of $P(k)$ would serve as a valuable baseline for the
interpretation of cosmological phenomena. Since the advent of
Inflation, cosmological structure formation theorists have been blessed 
with something rare in other fields of astrophysics, well motivated and
well specified initial conditions. Knowledge of $P(k)$ would add tremendous
extra power to quantitative studies of the formation of galaxies, clusters,
and other structures. 

One route to $P(k)$ uses observations of microwave background
anisotropies (the radiation counterpart
to the initial density fluctuations). However, estimates of the mass $P(k)$ 
derived from such measurements depend on the assumed values of the
cosmological parameters.
Furthermore,  the most accurate measurements of
microwave background anisotropies are presently confined to very large scales.
Much effort has therefore been spent on trying to infer $P(k)$ from surveys
of the galaxy distribution (see, e.g., Vogeley 1998 and references therein). 
Deriving an estimate of the primordial matter $P(k)$ from galaxy measurements
requires at the very least an understanding of how the present day
distribution of galaxies is related to the primordial distribution of mass.
This is essentially another definition of the commonly used term ``theory
of galaxy formation'', something which cosmology lacks at present  
in a quantitative enough form for this exercise to be carried out
(see, e.g., Kauffmann \etal 1998a). 
Even with such a theory, the complexity 
of the  processes involved, such as
gas dynamics, star formation, feedback, and non-linear 
gravitational collapse, promise to make it difficult to invert a theoretical
relationship to directly recover $P(k)$. 

Galaxies, however,  are not the only potential
probes of matter clustering. The \lya\ forest
seen in quasar spectra (Lynds 1971; Sargent et al.\ 1980) can also be used to
study mass fluctuations, but with two important differences. First, the 
framework  of standard cosmology  has provided us with a
well-motivated ``theory of \lya\ forest formation'', in which the bulk of
\lya\ absorption at high $z$ arises in a continuous, fluctuating, 
and highly ionized intergalactic medium 
(see, e.g., \cite{bi97}; \cite{hui97}; \cite{wkh98}; \cite{rauch98}
and references therein). Second, the  
situation described by the theory is simple, and leads to the prediction that 
an approximately local relationship holds between the absorbed flux 
in a QSO spectrum
and the underlying matter density, a 
relationship which can be inverted to learn about matter clustering. In 
particular, $P(k)$ itself can be recovered over a limited range of scales, 
as shown by Croft \etal (1998, hereafter CWKH). Here we will apply
the procedure of CWKH to recover $P(k)$ from a moderately large sample 
of QSO spectra of the \lya\ forest.

The modern picture of the \lya\ forest has arisen from 
theoretical studies of gas in the gravitational instability
scenario for the formation of structure. This theoretical picture was
originally proposed to explain observations of galaxy clustering and
formation. It was then discovered that, when the effect of a background UV
ionizing radiation field is included, the same theories naturally predict the
existence of QSO absorption phenomena. These predictions have been followed
using semi-analytic techniques (e.g., 
\cite{mcgill90}; 
\cite{bi93}; \cite{bi95}; \cite{reisenegger95};
\cite{bi97}; Hui \etal 1997), 
numerical simulations of cosmological hydrodynamics
(e.g., \cite{cen94}; \cite{zhang95}; \cite{hkwm96}; \cite{wb96};
\cite{theuns98}), and approximate N-body methods
(e.g., \cite{petitjean95}; \cite{gnedin98}).
The simulations and analytic models imply 
that the \lya\ forest arises primarily in diffuse gaseous structures of large
physical extent, consistent with the large transverse 
coherence length found in paired QSO observations 
(\cite{bechtold94}; \cite{dinshaw94}, 1995; \cite{crotts98}).
The absorbing structures that dominate the \lya\ opacity at high redshift
have gas densities fairly close to the 
cosmic mean, and they are still typically expanding with residual
Hubble flow, so
that the velocity width of absorption features seen in QSO spectra
corresponds mainly to a physical width (see the discussion in 
\cite{weinberg97}).
The effect of thermal broadening is minor, so that the
picture is qualitatively very different from previous representations of
the \lya\ forest features as discrete clouds
with a physical extent
much smaller than their thermal profiles. 

The physical state of the gas is largely governed by the competing processes
of photoionization heating by the UV background and adiabatic cooling due to
the expansion of the Universe. This places most of the
gas within a factor of 10 of the mean density on a power law 
temperature-density relation (Katz, Weinberg \& Hernquist
1996; \cite{hg97}) so that
\begin{equation}
T = T_{0} \rho_{b}^\alpha,
\label{eqn:td}
\end{equation}
where $\rho_{b}$ is the baryon overdensity in units of the cosmic mean.
The parameters $T_{0}$ and $\alpha$ depend on the reionization
history of the Universe and on the spectral shape of the UV background.
They are expected to lie in the ranges
$4000\;\K \la T_0 \la 15,000\;\K$ and  $0.3 \la \alpha \la 0.6$
(Hui \& Gnedin 1997).
In the moderate and low density
regions that produce the \lya\ forest, pressure gradients  
are small compared to gravitational forces, so that the
gas tends to trace the structure of the dark matter and
$\rho_{b} \simeq \rho$.
The optical depth for \lya\ absorption is proportional to
the neutral hydrogen density (Gunn \& Peterson 1965),
which  for this gas in photoionization
equilibrium is proportional to the density times the recombination rate. 
These proportionalities 
lead to a power law relationship between optical depth, $\tau$, and baryon
density, $\rho_{b}$:
\begin{eqnarray}
\tau &\propto & \rho_b^2 T^{-0.7} ~=~ A\rho_{b}^\beta, 
\label{eqn:tau} \\
A & = & 0.433
\left(\frac{1+z}{3.5}\right)^6 
\left(\frac{\Omega_b h^2}{0.02}\right)^2
\left(\frac{T_0}{6000\;{\rm K}}\right)^{-0.7} \;\times \nonumber \\
& & \left(\frac{h}{0.65}\right)^{-1}
\left(\frac{H(z)/H_0}{3.68}\right)^{-1} 
\left(\frac{\Gamma}{1.5\times 10^{-12}\;{\rm s}^{-1}}\right)^{-1}\; ,\nonumber
\end{eqnarray}
with $\beta \equiv 2 - 0.7\alpha$ in the range $1.6-1.8$.
Here $\Gamma$ is the HI photoionization rate, $H(z)$ is the
Hubble constant at redshift $z$, 
$h \equiv H_{0}/(100\;\kms\;{\rm Mpc}^{-1})$,
and $\rho_b$ is in units of the mean cosmic baryon density.
As representative fiducial values we have adopted the
baryon density $\Omega_b h^2$ advocated by Burles \& Tytler (1998), the
Hubble ratio $H(z)/H_0$ appropriate to an $\Omega_0=0.3$,
$\Lambda_0=0.7$ universe at $z=2.5$, the temperature $T_0$ for
mean density gas from the SPH simulation of Katz et al.\ (1996),
and the photoionization rate $\Gamma$ computed by 
Haardt \& Madau (1996) at $z \sim 2-3$.
Equation~(\ref{eqn:tau}) is based on a hydrogen recombination coefficient
$\alpha(T) = 4.2\times 10^{-13} (T/10^4\K)^{-0.7}$, which was adopted
by Rauch et al.\ (1997) as a good approximation to the recombination
coefficient of Abel et al.\ (1997) in the temperature range
that is most relevant for
the \lya\ forest.  Because equation (2) describes 
the analog of Gunn-Peterson absorption
for a non-uniform, photoionized medium (ignoring the effect of
peculiar velocities), 
we will refer to it as the Fluctuating Gunn-Peterson
Approximation (FGPA, see Rauch \etal 1997; CWKH;  Weinberg \etal 1998b).  
If we test the FGPA using artificial spectra extracted 
from simulations
(see, e.g., Figure 6 of Croft \etal 1997), we find that there is some
scatter in the relation between transmitted flux ($F=e^{-\tau}$)
and gas density because the spectrum is
measured in redshift space and because thermal broadening,
shock heating, collisional
ionization, and other effects included in the simulations are not accounted for
in the FGPA. However,  the regions that 
exhibit a substantial deviation from this approximation only constitute a small
fraction of the total length of the spectra. Any application of equation
(2) should be tested on a case by case basis with simulations. 
We do not make explicit use of equation (2) in the $P(k)$ recovery method,
but we will frequently refer to it to provide physical
motivation for our analysis.

The principle behind the $P(k)$ recovery method of CWKH is that the flux, $F$,
measured from QSO spectra constitutes a continuous, one-dimensional field 
whose relation on a point-by-point basis to the underlying matter
distribution is governed approximately by equation (2).
Applying a monotonic mapping of the
flux to give it a Gaussian probability distribution function converts a
spectrum to a line-of-sight initial
density field with arbitrary normalization. The one-dimensional power
spectrum of this density field can be inverted to give the three-dimensional
$P(k)$. The amplitude of $P(k)$ is set by running normalizing simulations
with different $P(k)$ amplitudes (assuming Gaussian initial conditions)
and picking the one for which the 
clustering of the flux in artificial spectra matches that in the observations.
The value of the uncertain parameter $A$ is determined in the normalizing
simulations by matching an independent observation, the effective
mean optical depth $\taueff \equiv -\ln\langle e^{-\tau}\rangle$.
It is this observational determination of $A$ that removes any
dependence of the derived $P(k)$ on unknown ``bias factors'' ---
the shape {\it and amplitude} of $P(k)$ are both recovered.

The rest of the paper is arranged as follows. The spectra of the \lya\
forest that constitute our observational data set are briefly
described in Section 2. The bulk of the paper (Section 3) deals
with the details of the $P(k)$ recovery, including tests of the
sensitivity of our results to continuum fitting and to the
resolution of the simulations used to derive the normalization of $P(k)$. 
In Section 4 we show that the artificial clustering that could be
caused by fluctuations in the UV ionizing background, which 
in principle could bias our
$P(k)$ measurement, is in practice too small to be significant on
the scales where we can measure $P(k)$. 
In Section 5 we present a tabulation of our results and
a power law fit to the data. We also compare our determination of
$P(k)$ with the predictions of specific Cold Dark Matter (CDM) models and
with recent measurements of galaxy clustering at $z = 3$ and $z =
0$. Finally, in Section 6 we summarize our main results and
outline directions for future work.
As Sections 2.2 through 4 focus on technical details of the application
of the CWKH procedure and
tests of its robustness, readers who are interested mainly in   
the final $P(k)$ result and a discussion of it should skip ahead 
to Section 5 after reading Section 2.1. 
A brief summary of the CWKH procedure is given at the beginning of Section 3.

\section{Observational data}

An advantage of studying the properties of matter clustering on relatively
large scales is that we do not necessarily need to use extremely
high resolution or high signal-to-noise ratio (S/N) data.
There will be a minimum scale below 
which the procedure for recovering the linear $P(k)$ does not
work because of the combined effects of
peculiar velocities, thermal broadening, and
non-linear evolution of the density field.
The tests of CWKH on hydrodynamic simulations showed recovery
of the correct linear $P(k)$ on large scales but 
suppression of power on small scales, which could be 
approximately modeled by smoothing 
the linear $P(k)$ with  a Gaussian filter,
of the form $\exp(-k^{2}r_{s}^{2}/2)$,
with $r_{s}=1.5/2\pi \hmpc$ 
($\sim 50 \kms$ at $z=3$).\footnote{Here we have included a factor of 
$2\pi$ that was omitted by error
from the formula in CWKH.  However, the tests with higher resolution
PM simulations in Section 4 below suggest that this cutoff scale
may have been partially set by the resolution of the CWKH simulations
(a point also made by \cite{haehnelt98}).} 
Information on smaller scales than this is therefore not directly useful to us
at present, and so we
can make effective  use of observations with spectral resolution (FWHM) as poor
as 2\AA, corresponding to 
a Gaussian dispersion $\sigma=0.7$\AA $=50 \kms$ at $z=2.5$.
The signal-to-noise ratio requirements 
are also not very stringent, basically because the \lya\ forest
data is in the form of a continuous  one-dimensional field, so that we 
 do not suffer from the shot noise present in galaxy data.
The errors that affect our determination of $P(k)$ are
mainly ``cosmic variance'' errors 
(more precisely, variations in the structure probed by a finite number
of spectra),
and the requirement is therefore
for a data set that samples as many independent sightlines as
possible. The signal-to-noise ratio and resolution do have a
secondary effect in that they determine the accuracy with which
the unabsorbed QSO continuum can be estimated. As explained in 
Section 3.1, uncertainties in this determination affect the
measurement of $P(k)$ on the largest scales.

\subsection{The QSO spectra}

The primary data sample used here represents a reasonable
compromise between the needs for resolution, signal-to-noise
ratio, and multiple sightlines. It is drawn from the survey of
Damped Ly$\alpha$ systems (hereafter DLA) by Pettini et al.
(1994, 1997) and consists of 19 QSO spectra obtained over the
period 1987 -- 1994 with the William Herschel telescope on La
Palma, Canary Islands and with the Anglo-Australian telescope at Siding
Spring Observatory, Australia. The spectra are reproduced in
Figure 1. The resolution ranges between 0.8~\AA\ and 2.3~\AA\
FWHM (typically $\sim 1.5$~\AA\ FWHM), and the signal-to-noise
ratio is between $\sim 10$ and $\sim 90$ (typically S/N $\geq 40$).
Further details of the data acquisition and reduction procedures
can be found in Pettini \etal (1997).

The QSO emission redshifts range from $z_{\rm em} = 3.23$ 
(Q0347$-$383) to $z_{\rm em} =  2.084$ (Q1331$+$170). The
spectra, which were designed to straddle the wavelength of
Ly$\alpha$ in intervening DLA systems, thus cover different
redshift ranges in the Ly$\alpha$ forest. In Figure 1, the
portion of each spectrum that was used in our analysis is shown
inside a solid box, which marks the wavelength region between the
QSO Ly$\alpha$ and Ly$\beta$ emission lines. It can be seen from
the Figure that most of our sightlines sample a redshift range
centered near $z=2.5$.

\begin{figure*}[t]
\centering
\vspace{12.5cm}
\includegraphics{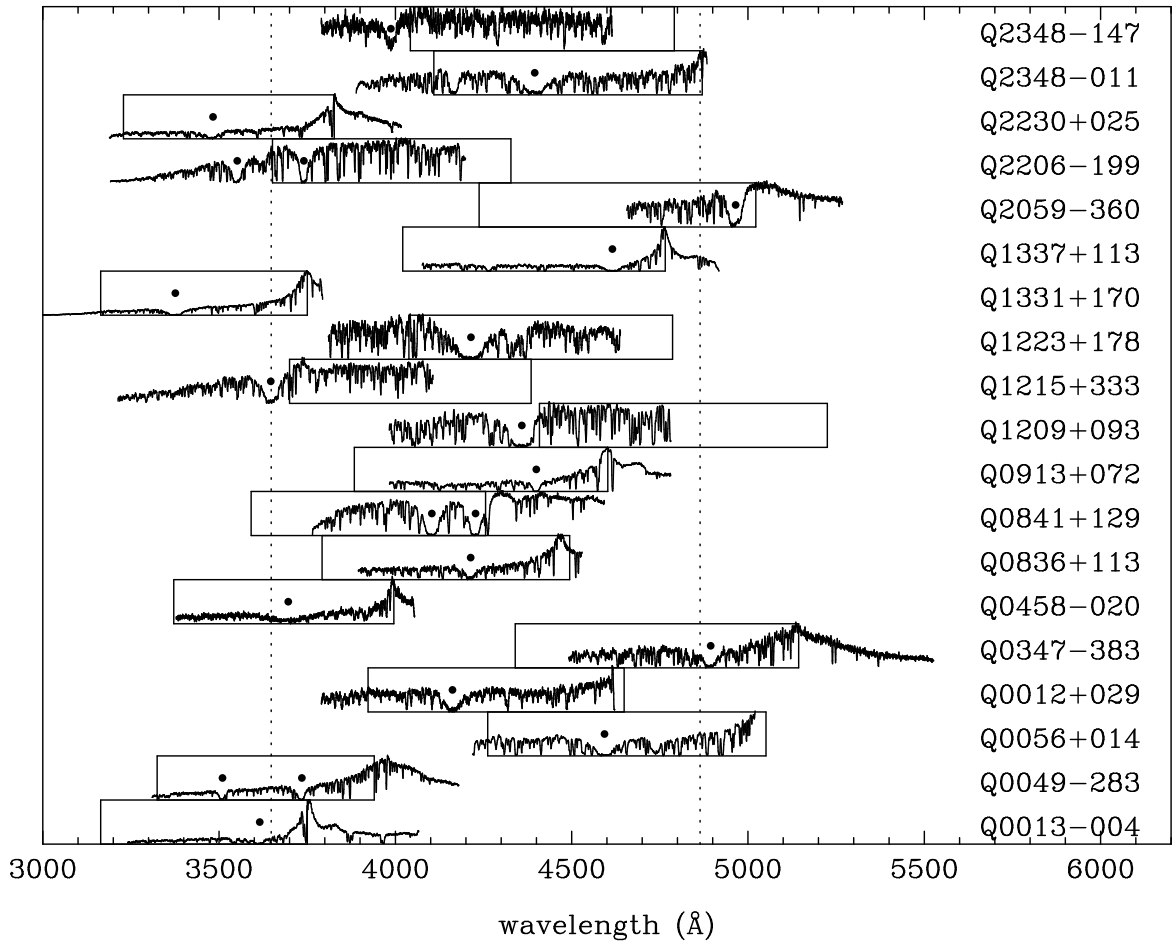}
\caption[spectra]{
The 19 QSO spectra that constitute the main data sample used in 
this paper. The solid boxes drawn around parts of
each spectrum represent the region from \lya\ to \lyb. Solid dots
are drawn at the redshifts of the DLA systems. The regions 
around these are excluded from the analysis (see text). Vertical dotted lines
are drawn at the wavelength of \lya\ at $z=2$ (left) and $z=3$ (right).
\label{spectra}}
\end{figure*}

For our analysis  we have constructed several subsamples of
the data from this primary sample of QSO spectra, as follows:

\noindent
(1) A ``fiducial'' sample, containing all the data between $z=2$ and $z=3$.
We will concentrate on this sample for most of our analysis. The restricted
range of $z$ is enforced so that the effects of redshift evolution are limited.
The total length of \lya\ to \lyb\ regions in this sample (once 
it has been prepared as described in Section 2.2 below)
is $4.8 \times 10^{5} \kms$, and the mean $z=2.5$.

\noindent
(2) The full sample, containing all the data. The total length is $6.4 
\times 10^{5} \kms$,
and the mean $z=2.4$. We will split this sample into 10 different subsamples
in order to estimate the errors on $P(k)$ (see Section 3.1).
  
\noindent
(3) A low-$z$ sample, for studying the effect of $z$ evolution, consisting 
of all the data with $z <2.4$. This sample has total length $3.2 \times 10^{5}
 \kms$ and mean $z=2.1$.

\noindent
(4) A high-$z$ sample, the data with $z> 2.4$, which has total length 
$3.2 \times 10^{5} \kms$ and mean $z=2.75$.

\noindent
The data preparation procedure (described in Section 2.2 below) 
removes regions around the DLA redshifts and near the QSO redshift
prior to analysis of any of these samples.

\begin{figure*}[t]
\centering
\vspace{8.0cm}
\includegraphics{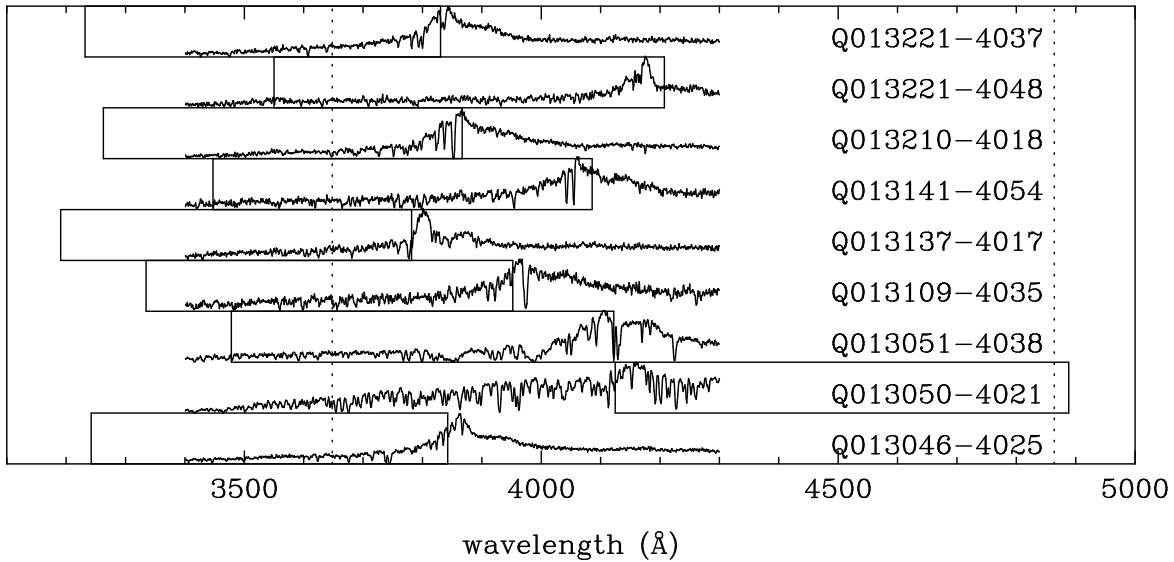}
\caption[spectra]{
The spectra of 9 QSOs in a field centered near 013145-403612.
$P(k)$ will be measured from this additional independent sample of data
and used to check our results from the main sample (see text).
 The solid boxes drawn around parts of
each spectrum represent the region from \lya\ to \lyb\ . Vertical dotted lines
are drawn at the wavelength of \lya\ at $z=2$ (left) and $z=3$ (right).
\label{spectrafield}}
\end{figure*}

In addition to analyzing these data, we make use of a secondary,
independent set of observations of the Ly$\alpha$ forest towards
nine QSOs in a southern field, 40 arcminutes in diameter, centered
at RA~=~01~31~45 and Dec~=~$-$40~36~12 (B1950). These data were
obtained in November 1986 by M. Pettini and R. Buss at the
Cassegrain focus of the Anglo-Australian telescope fed by the
FOCAP multi-fiber system, and are reproduced in Figure 2. All the
spectra cover the wavelength region 3400~--~4300~\AA\ with a
resolution of 2.2~\AA\ FWHM; the total exposure time of 56\,000~s
resulted in S/N $ \simeq 9 - 35$ (the QSO magnitudes range from 
$B=17.4$ to 20.7). The mean redshift of the useful portions of
these spectra is $z = 2.1$, conveniently the same as that of the
low-$z$ subsample of our primary data set described at point (3)
above. We therefore decided to analyze this secondary sample
separately, rather than combining it with the main data set, so as
to obtain an independent check on the results deduced from our
primary sample.

\subsection{Data preparation}

Before applying the $P(k)$ recovery machinery to the data, we need it to be in
the correct form, having been continuum-fitted. There are also a few more 
ways the data should be processed. We describe our data preparation
below and test the effects of varying the parameter choices in
Section 3.1.

First, we find the unabsorbed continuum level in the data in an 
automated fashion. We  use a standard iterative technique tested on 
simulations by Dav\'{e} \etal (1997) and  CWKH. The procedure is governed
by one free parameter, $L_{\rm fit}$, a length in \AA. We fit a third order
polynomial 
to a region in the QSO spectrum of length $2L_{\rm fit}$. We then discard all
points $2 \sigma$ below the fit line and fit again, iterating until 
convergence has been reached. The continuum level for the central
$L_{\rm fit}$ part of this region
is set by the final level of the polynomial. We then move 
$L_{\rm fit}/2$ onwards in wavelength and fit the next portion of the spectrum,
with the continuum fitted regions being joined together. We are therefore using
buffer zones of length $L_{\rm fit}/2$ around each  region.
The buffer zones stop the continuum from curving downwards artificially
if the $L_{\rm fit}$ region happens to end at a patch of high absorption.
For our fiducial sample, we use $L_{\rm fit}=50$ \AA.

Second, we prune the spectra to remove regions close to the QSO,
which might 
be affected by its ionizing radiation (the proximity effect, see, e.g., 
\cite{murdoch86}; Bajtlik, Duncan \& Ostriker 1988). 
We also remove the DLA systems, because 
in the  QSO spectra of the DLA survey they are obviously
present with a higher number density than the cosmic mean, which is
$0.2\pm0.05$ per unit $z$ interval at $z=2.5$ (Lanzetta \etal 1991).
They are also caused by gas of much higher density than we expect 
to be described by equation (2). One might worry that by pruning the
spectra we will somehow
bias the clustering in our sample, as we are excluding high density regions.
This is probably not the case, as high densities correspond
to saturated parts of the spectra and
tend to be given relatively low weight in the clustering analysis anyway.
It might also be thought that there is an opposite effect, whereby
the extra clustering in the mass around 
DLA systems could bias the overall clustering level upwards. This is also
extremely unlikely, as the DLA systems have a high enough 
space density that any enhanced clustering due to each one
can only extend over a tiny fraction of each
spectrum. In any case, when we carry out the $P(k)$ recovery,
we will test the effect of excluding a large
($100$ \AA) region around the DLA systems, and also of not excluding them
at all.

\begin{figure*}[b]
\centering
\vspace{8.0cm}
\includegraphics{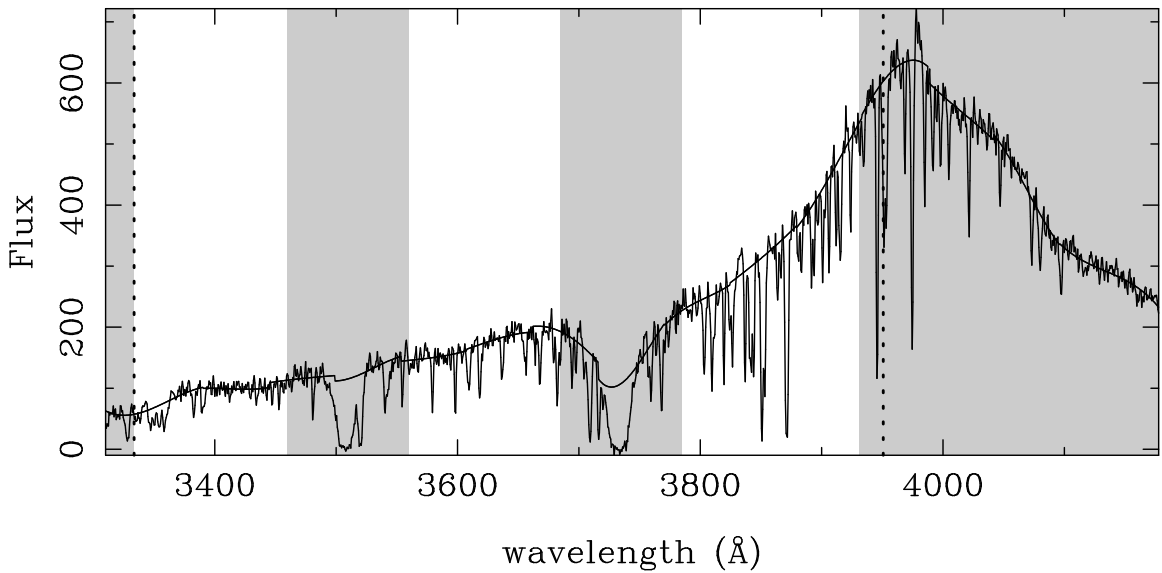}
\caption[contfit]{
An example QSO spectrum (Q0049-283). Vertical dotted lines
are drawn at the emission wavelength of \lya\ (right) and \lyb\ (left).
We have also plotted the continuum fitted by our procedure
(see text) with a fitting length $L_{\rm fit}= 50$ \AA. The shading denotes
regions excluded from the analysis. These are the regions blueward of  \lyb,
redward of  \lya, within $50$ \AA\ of either of the two DLA
systems, and within $20$ \AA\ of the QSO. 
\label{contfit}}
\end{figure*}

Third, we  attempt to mitigate the effects of evolution 
over the redshift range subtended by each individual sample. The most
noticeable effect of $z$ evolution is the decrease in the mean optical 
depth, which takes place as the Universe expands and the space density
of hydrogen atoms decreases. In an Einstein-de Sitter Universe, the
optical depth of photoionized gas evolves as $\tau \propto (1+z)^{4.5}$
owing to this effect. We follow Rauch \etal (1997)
and CWKH in rescaling the
fluxes in the spectra using this relation to the value they would have 
at the mean redshift of each sample (which is reasonable since all models
are approximately Einstein-de Sitter at these redshifts).
The spatial scales will also change due
to the expansion of the Universe. To first order, we can correct for this by
scaling all pixels to the size they would have in $\kms$ at the mean redshift
of the sample. In practice, this results in a constant scaling 
factor relating  pixel sizes in \AA\ to $\kms$. 
There will be additional, second order effects due to the 
change in $H$ over the 
redshift range, but these will be small, and model dependent, so we do not
attempt to correct for them.

Once we have treated the data as detailed above, we are left with
a number of disjoint spectrum segments, of various lengths, because of
the varying wavelength coverage and because the spectra have been broken up
by the exclusion of DLA systems. We discard all segments that are shorter
than a certain length (we use $100$ \AA), chosen to be at least a
factor of 3 larger than the maximum scale on which we measure $P(k)$,
so that the effects of convolution with the Fourier transform of the window
function are negligible. The data preparation procedure is illustrated in
Figure 3, which shows the continuum fit and excluded wavelength regions
for one of the spectra in the primary data sample.

\section{Recovery of the power spectrum}

The  method  we use for recovery of $P(k)$
from QSO \lya\ forest spectra is described and tested in detail
in CWKH. For completeness, we now give a 
brief account of the three principal steps in the procedure:

(1) We convert the spectra to one-dimensional linear density fields, 
by mapping the flux values in pixels monotonically to give them a Gaussian
probability distribution function (PDF)
with arbitrary normalization.
This ``Gaussianization''
procedure is motivated by the fact that gravitational
instability approximately preserves
the rank order of (smoothed) densities
(Weinberg 1992), so that one way of recovering the
initial density field is to monotonically map the final densities back to the
initial PDF, here assumed to be Gaussian. As the transformation between
flux and density given by the FGPA is also local and monotonic, mapping the
PDF of the flux directly to a Gaussian yields an initial density field,
to the extent that these approximations hold.
We note, however,
that our results for the shape of $P(k)$ are insensitive to the precise 
nature of the transformation applied to the observed flux. 
For example, it was found 
in CWKH that the power spectrum of the flux itself has the same
shape as the linear $P(k)$. We have found by numerical experiments
that any transformation 
of the density that suppresses the contribution of the high density 
regions (including Gaussianization, $F = e^{-A\rho^{\beta}}$, or even
truncation at $\rho/\bar{\rho}=5$) tends to produce a field whose power
spectrum has the linear $P(k)$ shape. 
Thus, Gaussianization is not indispensable to the $P(k)$ recovery
method, although it appears to be a useful way of ``regularizing''
spectra and thus reducing noise in the recovery (CWKH).
In Section 5.4, we will briefly compare the shape of the primordial 
$P(k)$ to the non-linear $P(k)$ of the mass in simulations.

(2) We measure $\p1dk$, the one-dimensional power spectrum of this
density field, using a Fast Fourier Transform. We  convert this 
$\p1dk$ to the 
three-dimensional $P(k)$ by differentiation
(Kaiser \& Peacock 1991; CWKH),
\begin{equation}
P(k)=-\frac{2\pi}{k} \frac{d}{dk}\p1dk.
\label{eqn:invert}
\end{equation}
Equation (3)
assumes that the distribution of matter is isotropic with respect to the line
of sight. Redshift-space
distortions caused by peculiar velocities
mean that this is not strictly true (Kaiser 1987), and these distortions
must be taken into account for a truly accurate inversion of one-dimensional
clustering (Hui 1998). We find in simulation
tests (e.g., those in CWKH) that any error in the shape of the 3D $P(k)$ caused 
by redshift-space distortions is small and well within the statistical
errors for the present observational determination of $P(k)$,
although it could have a noticeable effect in some future samples.
In step (3) below, we use our measured $P(k)$ shape as an input to
the normalizing simulations.
The $P(k)$ that we use for this purpose corresponds
to $P(k)$ from equation (3) multiplied by $\exp(k^{2}r_{s}^{2}/2)$,
with $r_s=34 \kms$, in order to compensate for power lost on small scales
due to the finite resolution of the observations, as discussed
in Section 2.  However, we will only compare our recovered
$P(k)$ to theoretical predictions 
on scales where $k < 0.5/r_s$. In CWKH it was shown that the recovered
$P(k)$ on these larger scales is insensitive to the details of the power
restoration on smaller scales.

(3) The $P(k)$ resulting from step (2) is still of arbitrary amplitude. 
To determine the normalization, we use simulations that have Gaussian 
initial conditions (i.e., random Fourier phases) and an initial power
spectrum with the same shape as our measured $P(k)$ (with small scale
power restored as explained above) but with various linear theory
amplitudes. The higher the power spectrum amplitude, the larger
the fluctuations in the evolved mass density field, and hence the
larger the predicted fluctuations in the observed flux.  
We can therefore 
 pick the correct $P(k)$
amplitude by comparing clustering in spectra 
extracted from these simulations with the observations themselves.
The statistic that we choose to make the comparison
with is the three-dimensional power spectrum of the flux 
(more precisely, the power spectrum of $F/\langle F\rangle-1$, where $F$ is
the ratio of the observed flux to the unabsorbed continuum).
To distinguish this
from $P(k)$ of the mass in plots, we will plot 
\begin{equation}
\deltaf=k^{3}P_{F}(k),
\end{equation}
 where $P_{F}(k)$ is the 
three-dimensional power spectrum of flux. 
The quantity $\deltaf/2\pi^2$  is the contribution
to the variance of the flux from an interval $d\ln k =1$.
We run the simulations using the
PM approximation, where we use a standard PM N-body code
to  evolve the mass distribution and assume (a) that the 
gas pressure effects in the low and moderate density regions are unimportant,
so that the gas traces the dark matter, and (b)  that the gas follows the
power law temperature-density relation discussed in Section 1.
The CWKH tests show that the PM approximation gives accurate 
predictions of the flux power spectrum relative to full hydrodynamic
simulations.
In making the artificial spectra from the normalizing simulations, there is
one free parameter, in addition to the $P(k)$ amplitude, 
that can also influence the
amplitude of flux fluctuations: the parameter $A$ of
equation (2). Although it depends on physical quantities that are not known
individually (such as $\Omega_{b}$, $\Gamma$, and $T_{0}$),
$A$ as a whole can be set by appealing to one observational
measurement that we  have not yet used, the mean flux level in the spectra.
We therefore fix $A$ in our normalizing simulations by picking the value for
which the spectra have the same mean flux level as the observational
measurements of Press, Rybicki \& Schneider (1993,
hereafter PRS). The mean flux level, $\langle F \rangle$, 
is often expressed in terms
of an effective optical depth, $\taueff=-\ln{\langle F \rangle}$.
Any uncertainty in the value of $\taueff$,
and hence in $A$, is directly linked to uncertainty in the amplitude
of $P(k)$. Our choice of a particular observational determination of this 
quantity could therefore affect our results appreciably. We will discuss this
issue further later in the paper. 

Given that the above procedure seems rather complicated, one could ask why 
we do not simply attempt a direct inversion of the flux to a mass
distribution, using the FGPA as a guide,  along
the lines of the procedure proposed recently by Nusser \& Haehnelt (1998).
For purposes of determining the primordial $P(k)$, our more
indirect procedure is more robust and more broadly applicable,
for several reasons. First, our approach is 
relatively insensitive to what is
occurring in saturated regions, which cannot be inverted directly 
from  observations of \lya\ absorption alone, and which in any case
are less likely to obey
the FGPA. Second, our method relies mainly on
large scale clustering information;
it therefore does not require data that fully resolve all \lya\ features.
Finally, the use of simulations in the normalizing procedure
provides a convenient way to estimate the unknown
parameter $A$, and it automatically includes the
effects of non-linear gravitational evolution and peculiar
velocities.

In our method of deriving $P(k)$, the assumption that primordial 
fluctuations are Gaussian enters mainly into the normalization step (3),
since we use Gaussian fluctuations to initialize our normalizing simulations.
If we adopted non-Gaussian initial conditions with the same $P(k)$ shape,
then the $P(k)$ amplitude required in order to match the observed flux
power spectrum with the observed $\taueff$ as a constraint might be
different.  The Gaussian assumption also motivates
the Gaussianization procedure applied in step (1), but since the
derived shape of $P(k)$ would be similar even without Gaussianization,
it seems likely that recovery of the shape of $P(k)$ does not depend
much on the assumption of Gaussian initial conditions.
However, all of the tests in CWKH and in this paper are for initially
Gaussian models, and the success of our method in recovering the
shape and amplitude of $P(k)$ in non-Gaussian models would need to
be tested on a case-by-case basis.

\subsection{The shape of $P(k)$}

\begin{figure*}[t]
\centering
\vspace{7.7cm}
\includegraphics{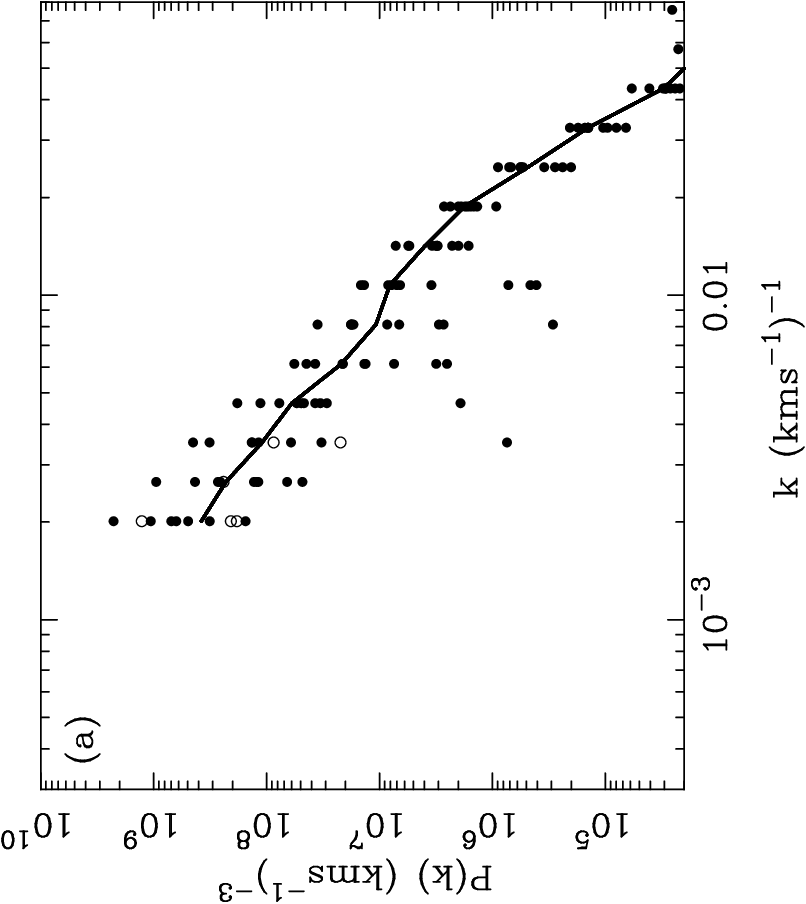}
\includegraphics{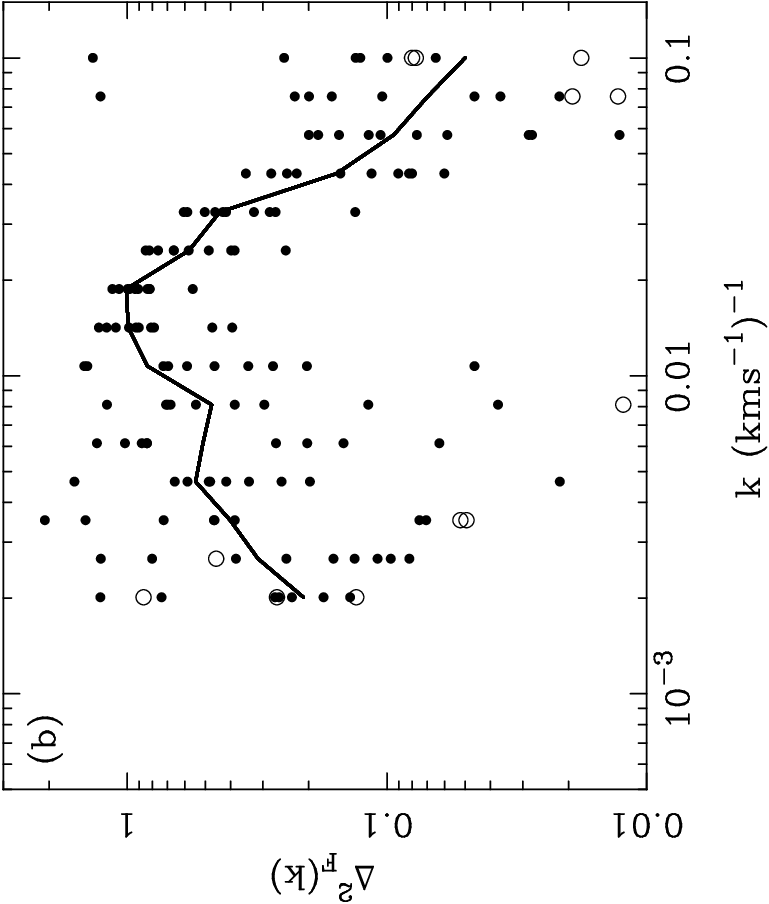}
\caption[10bits]{
(a) $P(k)$ for the Gaussianized flux. The solid line is measured
from the spectral
regions between $z=2$ and $z=3$ (the fiducial sample).
The dots were made by dividing
the full sample into ten separate pieces and calculating $P(k)$ individually
for each one.
Open circles are plotted at the modulus of negative data points.
(b) The 3D flux power spectrum, $\deltaf \equiv k^3 P_F(k)$,
for the same samples plotted in (a). 
\label{10bits}}
\end{figure*}

We now turn to the analysis of the observational data. First, we measure the
shape of $P(k)$ as described in steps (1) and (2) above, and also 
measure $\deltaf$. In order to estimate errors, we take the 
whole data set [sample (2) of Section 2.1] and split it into 10 subsamples,
of roughly equal length. We estimate $P(k)$ and $\deltaf$ individually
for each of the subsamples; the results are plotted as points in 
Figure \ref{10bits}.
When Gaussianizing the flux to yield an initial density field,
we set the $\sigma$ of the Gaussian
PDF to be the same for each of the subsamples. 
For several of the subsamples, the values of
$P(k)$ and $\deltaf$ for the largest scale 
plotted in Figure \ref{10bits}a are unphysically negative,
as are a few measurements on smaller scales. 
This can occur when the measured one-dimensional power spectrum
is noisy, as the noise may result in regions with a positive slope,
$\frac{d}{dk}\p1dk$, so that the inversion of equation (3)
yields negative values for $P(k)$.
The largest scale point plotted marks the limit where
cosmic variance noise is small enough for this sample 
to allow us to make a reasonable inversion
from one-dimensional to three-dimensional clustering. We will see later that
the real maximum scale on which 
we can believe the $P(k)$ measurement appears to
be slightly smaller, and is set by continuum fitting.

The solid lines in Figure \ref{10bits} are the $P(k)$ and
$\deltaf$ measurements
from the fiducial sample [sample (1) of Section 2.1], on which we will 
base most of our analysis.
We will assign error bars to these measurements that are derived from the 
fractional error in the mean of the measurements from the 10 subsamples of the 
full data set described above. We base our error estimate on the variance
among subsamples of the {\it full} data set rather than 
the smaller, $z$-limited, 
fiducial data set for two reasons: the inversion from 
1D to 3D clustering is more manageable with the larger subsamples, and
the errors based on a data set with a larger range in $z$ should
be conservative, as there will be extra variance  introduced by
the larger $z$ evolution. We therefore assign fractional errors from the
full sample to other samples, allowing for the difference in the
number of independent data elements in each sample by
scaling the fractional errors
by the ratio of the square roots of the sample lengths.  

We can see from Figure \ref{10bits}
that there is significant variation between the results
for the different subsamples. Each subsample corresponds
to roughly  the length of
one full \lya\ to \lyb\ region in a spectrum. The errors increase towards
large scales because we are averaging over fewer independent modes. On
small scales, we see a  turnover, due to the finite observational
resolution of our data sample. The lowest resolution 
data that forms part of our sample has a FWHM resolution of 2.3 \AA. 
As discussed in Section 2, this is similar to the smallest scale for
which the simulation tests of CWKH verified that the linear
$P(k)$ can be correctly recovered.
Because of this limitation, we should only regard our results
on scales larger than $k \sim 0.02 \invkms$ as being representative
of the true shape of $P(k)$.

\begin{figure*}[b]
\centering
\vspace{7.7cm}
\includegraphics{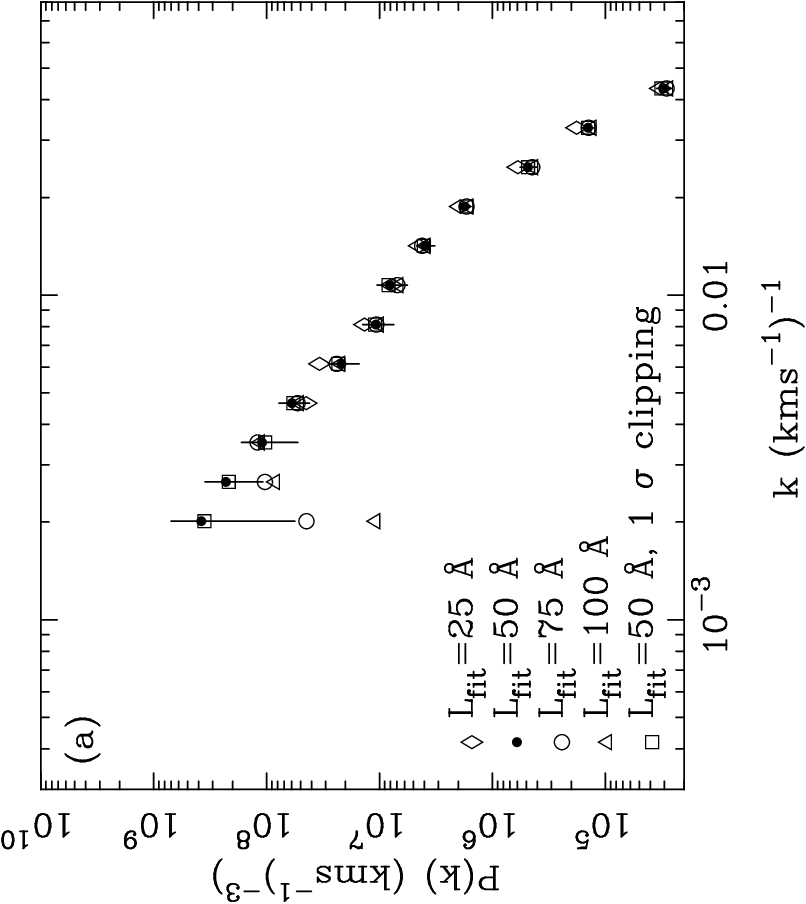}
\includegraphics{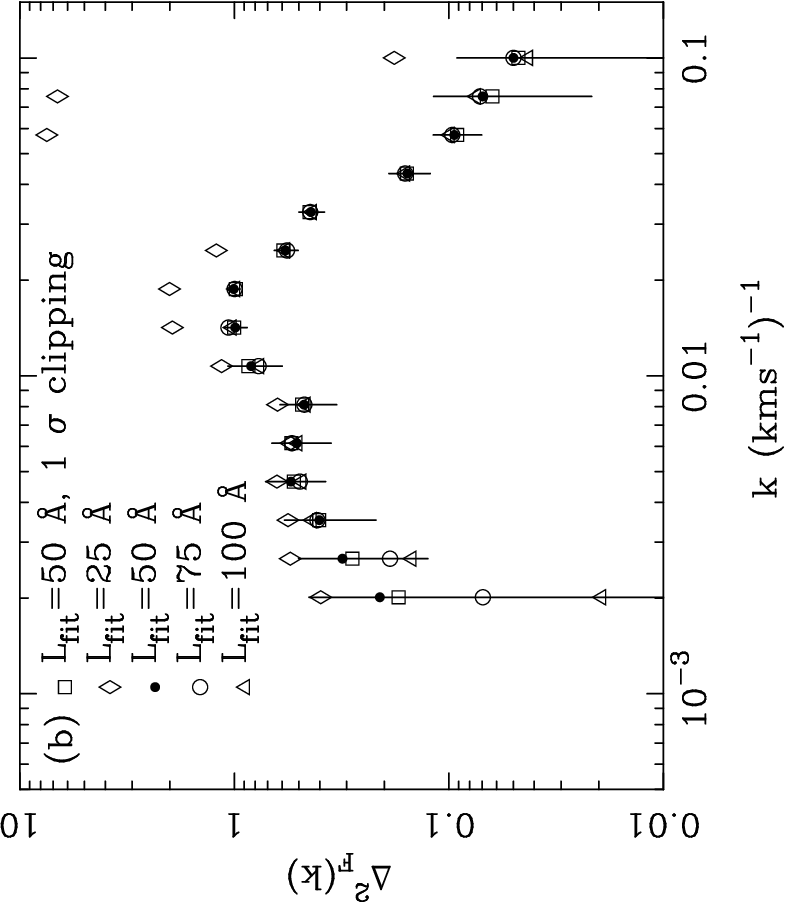}
\caption[junk]{
Tests of the effect of changing the length over which the continuum 
is fitted, $L_{\rm fit}$. See Section 2.2
for details of the fitting procedure.
(a) The Gaussianized $P(k)$. (b) The 3D flux power spectrum, 
$\deltaf$.
Error bars are derived from
the error on the mean taken from splitting the sample
into 10 subsamples.
\label{contfitpk}}
\end{figure*}

The data preparation procedure described in Section 2.2 involves several
operational parameters, the choice 
of which could conceivably affect
our results. One of these is the length $L_{\rm fit}$ over which the continuum
is fitted. In Figure \ref{contfitpk} we test 
the effect of using different values of
$L_{\rm fit}$. Again we plot both $P(k)$ and $\deltaf$, this time for the fiducial
sample. The error bars have been determined in the manner
explained previously. 

The smallest $L_{\rm fit}$ we try, $25$ \AA,
is obviously too small, being similar in size to the 
largest wavelength on which we measure $P(k)$. 
We try this value as an experiment, 
to see how this poor choice of $L_{\rm fit}$  will affect our measurements.
By comparing panels (a) and (b) of Figure \ref{contfitpk},
we can see that the effects
are different for $P(k)$ and for $\deltaf$. On  scales
$k < 4 \times 10^{-3} (\kms)^{-1}$ and larger, the
continuum fitting has completely eliminated power in $P(k)$, but $\deltaf$
has increased. This difference may reflect
the fact that the Gaussianized field
used to measure $P(k)$ has more prominent low density regions,
as the Gaussianization stretches out the PDF of low densities into a Gaussian
tail. These low density regions, being closer to the continuum, may be more
influenced by fitting. When we use more reasonable values for $L_{\rm fit}$,
including our fiducial value of $50$ \AA, we can see that the two
largest scale points are affected by the choice of fitting length, and
for $k < 2 \times 10^{-3}\invkms $
the systematic variation is outside the statistical 
errors. We will therefore discard the
largest scale point when making use of our results. It is interesting
that increasing $L_{\rm fit}$ appears to yield less power, although it is not
certain whether this represents a trend or merely a statistical fluctuation.

One might worry that with our relatively low spectral
resolution we will fit the continuum systematically low everywhere.
One way of checking to see if this is a problem is to change the 
clipping level below which points are discarded during the fitting process.
The usual value is $2 \sigma$ (where $\sigma$ is the error
on the flux at a point), but if we change it to $1 \sigma$ we tend to
fit the continuum much higher, almost certainly too high. The mean effective
optical depth of the sample, $\taueff$,
increases by $25 \%$ to 0.30,
but this change has no direct impact because we use the PRS
measurement of $\taueff$ to fix the value of $A$, not the
value from the sample itself.
What is important is that 
$P(k)$ and $\deltaf$ hardly change at all,
as shown by the open squares in Figure \ref{contfitpk}.
We have also tried raising the continuum uniformly everywhere by $10 \%$, 
and we again find that this has a negligible effect on our results.

As we will see by considering other potential factors,
the continuum fitting process appears to act as a limit to the largest
scales on which we can measure $P(k)$
from the current data set.  It may be that data with higher
spectral resolution would allow more accurate continuum
fitting and hence enable measurement of $P(k)$ at larger scales.
In future work we plan to carry out a systematic analysis
of continuum fitting procedures using larger volume
simulations (for which the true continuum is known).
Such analysis might suggest better ways of determining the continuum,
perhaps involving a totally different technique
(see, e.g., PRS), and it would help us understand the interplay
of spectral resolution, signal-to-noise ratio, and continuum
fitting in limiting the accuracy and dynamic range of $P(k)$ 
recovery.  Continuum fitting also has an important impact on 
other statistical measurements of the \lya\ forest, such as
$\taueff$ and the flux decrement distribution function
(\cite{rauch97}).  Although a detailed investigation of these
issues is beyond the scope of
this paper, we can already surmise from consideration of
Figure \ref{contfitpk} that our $P(k)$ measurement is likely
to be reliable out to a wavelength $\lambda \sim 2300 \kms$, which for
$\Omega_{0}=1$ corresponds to a comoving scale
$\sim 12 \hmpc$. On scales smaller than this,
reasonable variations in the continuum fitting procedure have
no significant influence on our derived $P(k)$.

\begin{figure*}[t]
\centering
\vspace{7.7cm}
\includegraphics{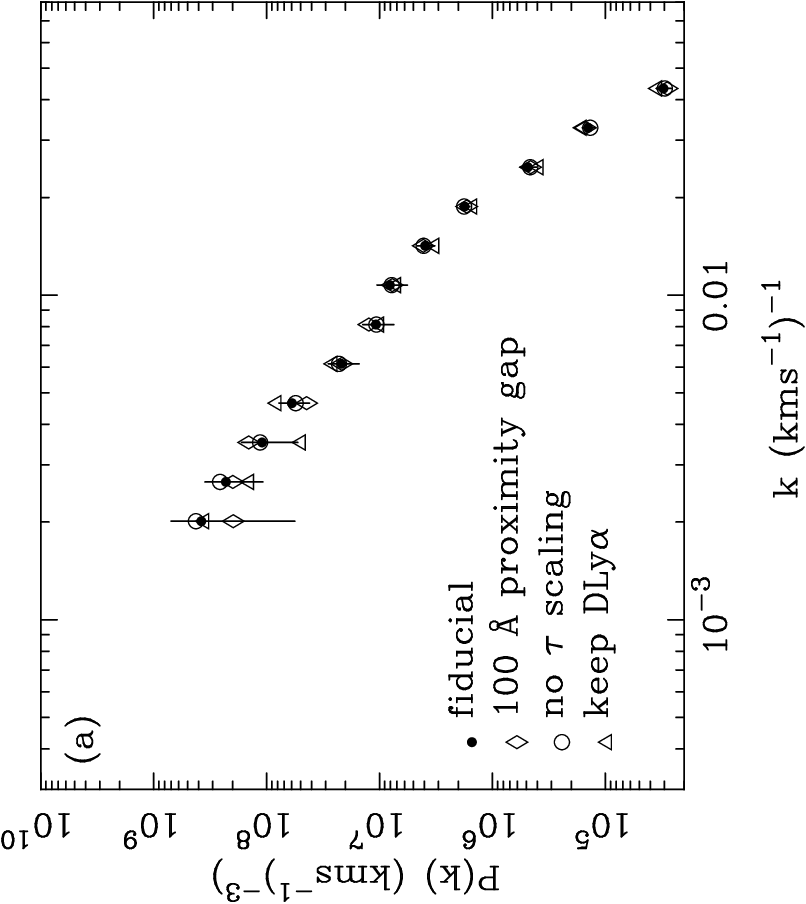}
\includegraphics{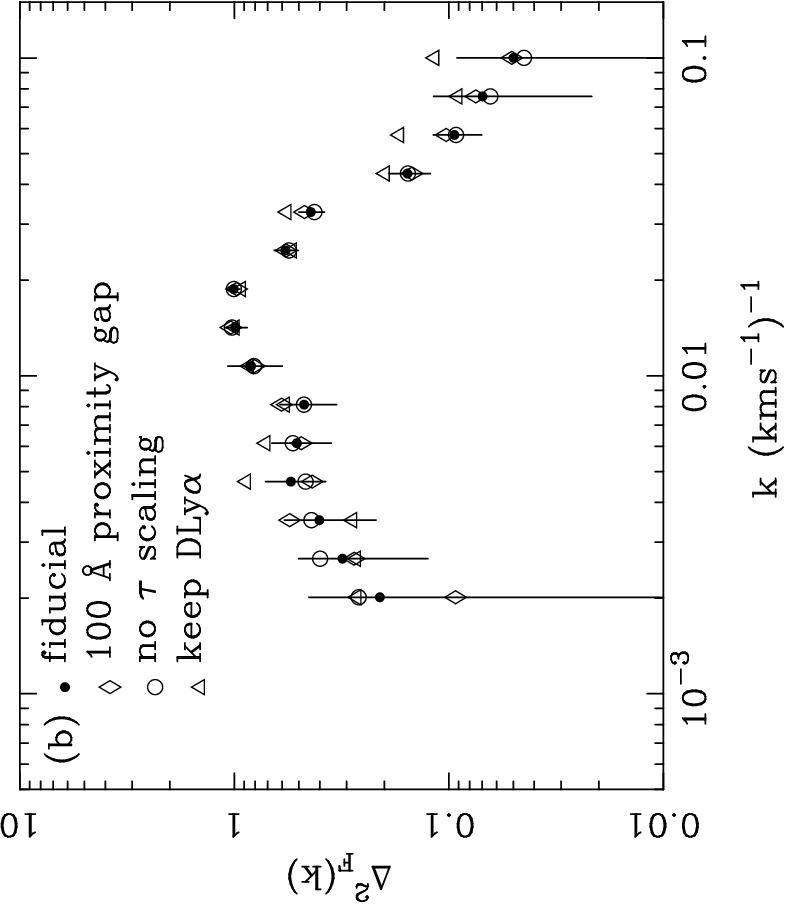}
\caption[junk]{
Tests of the data preparation  procedure. The points labeled ``fiducial''
are for $z=2$ to $z=3$ and have the continuum fitted over a 50\AA\ region, 
have $(1+z)^{4.5}$ 
scaling of $\tau$, a 20 \AA\ proximity gap, and a 100 \AA\ region excluded 
around DLA systems. The other points show the effect of
varying these parameters.
(a) The Gaussianized $P(k)$. (b) The 3D flux power spectrum, $\deltaf$.
\label{vary}}
\end{figure*}

In Figure \ref{vary} we vary the other 
parameters used in the data preparation of Section 2.2
and compare with results for the fiducial parameter choices.
We can see that changes such as not scaling the optical 
depths to the mean redshift, or increasing the QSO proximity gap
to $100$ \AA\ from the fiducial value of $20$ \AA, cause only minor
changes to $P(k)$, within the $1 \sigma$ errors. Even keeping the DLA systems
as part of the analyzed portion of the spectra,  obviously not a
reasonable thing to do, does not change the results by much; there is a small
change in $P(k)$ and $\deltaf$ near
$k\sim 0.05 (\kms)^{-1}$, which 
is probably the signature of power on the scale of the DLA systems themselves.
These tests therefore increase our confidence in the robustness
of the measurements of  $P(k)$ and $\deltaf$. 
For larger, future samples, different treatments of the data may
yield systematic differences of results that 
rival the statistical errors. This does not appear to be the 
case here, indicating that the procedures we have adopted
are adequate for our current data and objectives.

\subsection{The normalization of $P(k)$}

As outlined previously, we use simulations to normalize our estimate of
$P(k)$. The $P(k)$ we use as an input to the normalizing simulations
is slightly different from  the 
measured $P(k)$  plotted in Figures \ref{contfitpk} and \ref{vary},
in that we restore small scale power 
that was suppressed by the limited observational
resolution. In CWKH, it was shown that missing 
power on small scales only has a small effect on $\deltaf$ at the large
scales we use for normalization (see Figure 9 of CWKH). 
We therefore do not need to make this correction for lost small scale power 
very precisely. As described at the beginning of
Section 3, we ``unsmooth'' $P(k)$ using a Gaussian filter, so that 
\begin{equation}
P_{S}(k)=P(k) \times e^{k^2 r_s^2/2},
\end{equation}
where $r_s=34\kms$ and $P_{S}(k)$ is the power spectrum used
in the normalizing simulations. 
We also extrapolate $P(k)$ above the largest measured
point using an $n=-1$ power law.

The PM simulations have $128^{3}$ particles on a $256^{3}$ grid in
a periodic box $4170 \kms$ on a side. 
We assume $\Omega_{0}=1$ and $\Lambda_{0}=0$
(and $H_{0}=50 \kmsmpc$) when
running the simulations, so that 
the box size is $22.22 \hmpc$ comoving. CWKH have shown that the choice of
cosmological parameters has a negligible effect on the results 
provided that one works in the observed $\kms$ units.
We choose $\Omega_{0}=1$ for the simplifying reason 
that we can use different outputs of a single simulation to
represent different mass fluctuation amplitudes, since $\Omega=1$ at all
redshifts and the linear theory fluctuation amplitude is
proportional to the expansion factor, $a(t)$.
The initial density field is set up using $P_{S}(k)$ and Gaussian random 
phases. We average results from 4 different realizations that use
different random seeds. The simulations are run so that the expansion
factor, $a$, increases by a factor of 16.8 from the initial
conditions to the most evolved output, in 84 equal
steps of $\Delta a=0.2$.

\begin{figure*}[b]
\centering
\vspace{8.7cm}
\includegraphics{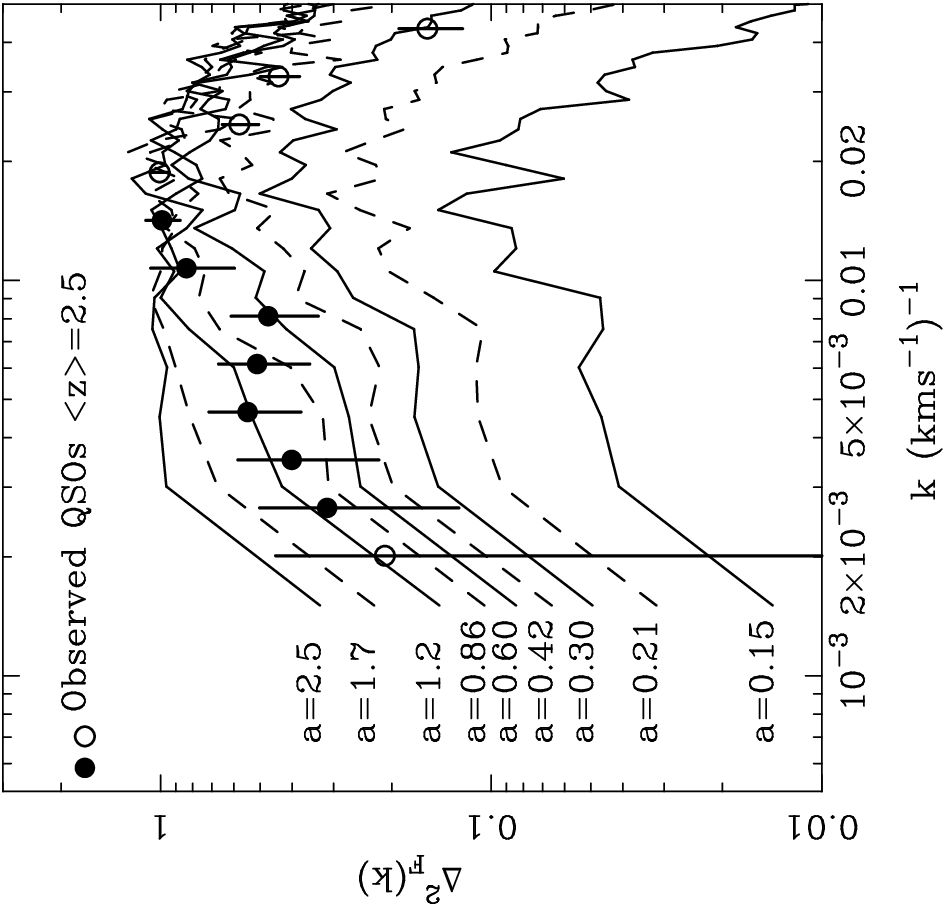}
\caption[junk]{
The 3D power spectrum of the flux for the fiducial sample, compared
to outputs from normalizing simulations that use 
$128^{3}$ particles in a $22.22 \hmpc$ box ($\Omega_{0}=1$).
The different curves are labeled by the expansion
factor at each simulation output, normalized so that $a=1$
corresponds to the amplitude determined to best fit the observational results. 
Only the solid points are used to fix the normalization.
The open points on smaller scales are affected by the finite resolution
of the spectra and the simulations, and largest scale point is
systematically uncertain because of continuum fitting (Section 3.1).
\label{norm}}
\end{figure*}

We extract spectra from the simulations, for several 
different output times, using the methods described in 
CWKH. We use a temperature-density relation of the 
form given by equation (1), with $T_{0}=5600$K and $\alpha=0.6$.
We adjust the mean effective optical depth
$\taueff$ (by varying $\Omega_{b}^{2}/\Gamma$) so that 
$\taueff=0.28$, the PRS value at $z=2.5$. 
We extract 2000 spectra in total at each output time and calculate $\deltaf$
from them. The results are plotted in Figure \ref{norm},
where the curves for different
output times are labeled with the expansion factor $a$, and $a=1$ has been
chosen to correspond to the normalization appropriate for the observational
results (which we shall describe below). From Figure \ref{norm}, we can see
that changing the underlying amplitude of mass fluctuations results
in a clear change in $\deltaf$ (recall that $\taueff$ is the same for
all spectra). We will restrict our quantitative use of the 
data to large scales, with $k < 0.02 \invkms$, 
which in practice means using the
2nd through the 8th observational points (we discard the first point because
of continuum fitting uncertainties). On smaller scales
we do not know the initial  $P(k)$ accurately, and the predicted
$\deltaf$ depends on  physical assumptions
and the resolution of the simulations.
 
To determine the normalization of $P(k)$,
we must decide which of the
$\deltaf$ curves in Figure \ref{norm} (or interpolation between these curves) 
is closest to the observational results. 
One possible method for determining the correct normalization
involves a maximum likelihood fit of the simulation results
to the observational data, where we seek to maximize the likelihood
by minimizing 
\begin{equation}
\chi^{2}=\sum_{ij} [\Delta^{2}_{F}(k_{i})-\Delta^{2}_{Fsim}(k_{i},a)]
C^{-1}_{ij} [\Delta^{2}_{F}(k_{j})-\Delta^{2}_{Fsim}(k_{j},a)].
\end{equation}
Here $\Delta^{2}_{F}(k_{i})$ is the observed value of
$\deltaf$ at $k=k_{i}$ and $\Delta^{2}_{Fsim}(k_{i},a)$
is the equivalent quantity measured from simulation outputs at expansion
factor $a$. The covariance matrix of the data points, $C$, is estimated 
using the observational data split into subsamples.
If we use this procedure, we obtain a normalization of $P(k)$ with errors of 
$(+10\%,-9.5\%)$.  We have also applied this procedure using
a jackknife estimator
to determine $C$ from the ten subsamples, with very similar results.

Despite its statistical logic, we have decided not to adopt the
maximum likelihood determination of the normalization and error
but instead to rely on a simpler estimator.
There are two main reasons for this. First, we find that the above
procedure yields an unrealistically low value of $\chi^{2}$ 
for the best fitting output. 
The low $\chi^2$ arises because the $P(k)$ used in the simulations
is measured from the observational data themselves, so that
the {\it shape} of $\deltaf$ in the simulations is more correlated
with the observations than the cosmic variance error bars suggest.
Second, and probably more important, the points on small scales, with small
statistical errors, are weighted most highly in the maximum likelihood fit.
These points are also those most likely to be affected by systematic 
errors in the simulations resulting from resolution effects (see below)
or uncertainty in the input physical assumptions (see CWKH).
We therefore adopt an estimator that depends more evenly on
the data points and that condenses the information about the
amplitude of $P(k)$ into one number,
\begin{equation}
S=\sum_{i}\Delta_{F}^{2}(k_{i}).
\end{equation}
The sum is over the 2nd to the 8th data points, as explained above.
Because the variance of the flux is 
$\int_0^\infty \Delta_F^2(k) d{\rm ln}k/2\pi^2$ and our
data points are evenly spaced in ${\rm ln}k$, 
the quantity $S$ is simply the contribution to the flux
variance from the wavenumber range over which we estimate
the power spectrum.
Although this estimator does not weight the data optimally in a 
strictly statistical sense, it is less sensitive to systematic
errors on small scales, and by using it we arrive at a conservative
estimate of the $P(k)$ normalization error.

The observed value of $S$ is our diagnostic for the amplitude of
$P(k)$ --- the higher the amplitude, the larger the value of $S$
predicted by the normalizing simulations.  We choose the
best fit amplitude to be the one for which the predicted $S$ matches
the observed value (using linear interpolation between the two closest
simulation outputs).  We obtain the $1\sigma$ uncertainty by measuring 
$S$ separately for each of the 10 subsamples
of the full data set [sample (2) in Section 2.2],
converting the $1\sigma$ error on the mean of $S$ into 
a corresponding uncertainty in the mass fluctuation amplitude.
Since the relationship between $S$ and the amplitude is fairly linear,
we would get similar results if we instead determined the amplitude
separately for each subsample and took the $1\sigma$ error
on the mean amplitude.  Normalization
errors for subsets of the
data, such as the fiducial sample and the other samples of Section 2.2,
come from scaling the errors on $S$ by the ratio of the 
square roots of the lengths of the spectra involved,
as was done with the errors on the individual $P(k)$ points.

After applying this
procedure, we find the $\pm 1 \sigma$
uncertainty on the normalization of the fiducial ($z=2-3$) 
sample to be $(+17.0\%, -16.5\%)$
in the fluctuation amplitude $a$. The normalization itself
is $15 \%$ higher in $a$ than that which results from applying the
maximum likelihood fit of equation (7).

\begin{figure*}[t]
\centering
\vspace{7.7cm}
\includegraphics{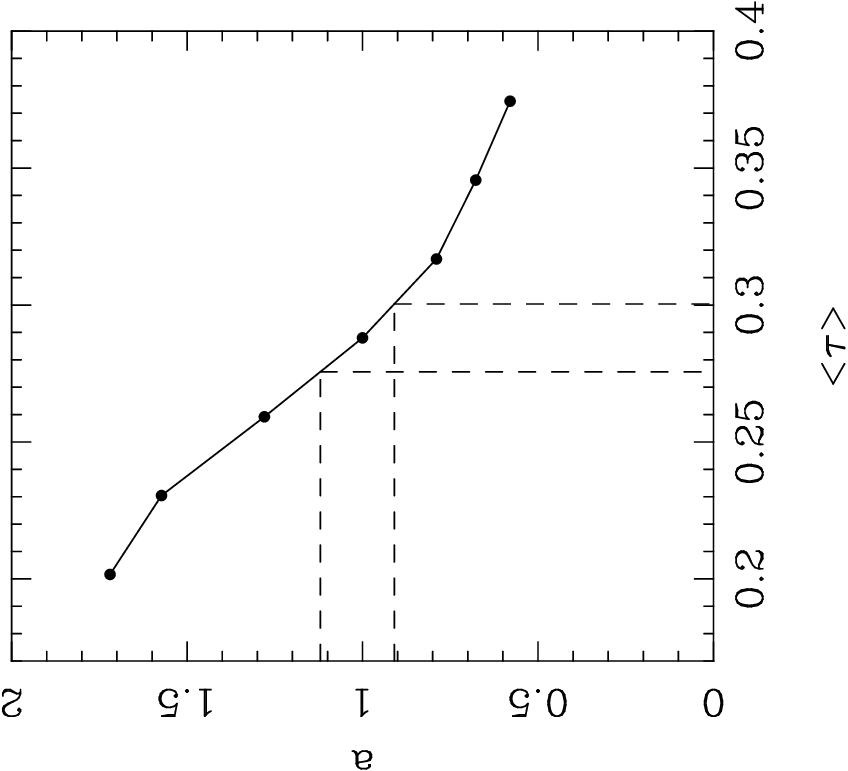}
\caption[junk]{
The effect of varying $\taueff$ in the normalizing simulations.
The dashed lines show the $1\sigma$ errors on the value of $\taueff$
from PRS and how they correspond
to an uncertainty in $a$. The errors are $\sim\pm4\%$ on $\taueff$
and ($+12\%, -9\%$) on $a$. 
\label{taubar}}
\end{figure*}

An important additional source of error is uncertainty
in the value of $\taueff$. 
The value we use is 
given by the PRS formula $\taueff=0.0037(1+z)^{3.46}$. 
PRS measured their result from a sample of 29 low resolution QSO spectra.
They estimated the continuum in the \lya\ forest region by extrapolating
the continuum observed on the red side of the \lya\ emission line.
The results are consistent with  those measured from  high resolution 
Keck spectra (Rauch \etal 1997) using a  
polynomial continuum fitting technique blueward of \lya. 
The smaller Keck sample has larger statistical errors, but its consistency
with the PRS result makes us reasonably
confident that the value of $\taueff$ we use is close to the true one.
However, we note that discrepant, lower results for $\taueff$
have been published by other authors (e.g., \cite{zl93}; 
\cite{dobrzycki96}), and the issue is not settled.

We quantify the influence of $\taueff$ on the $P(k)$ amplitude
by making new spectra from our
normalizing simulations, with different values of $\taueff$. We then 
carry out our normalizing procedure using the weighted sum of equation (7)
to find the best fitting value of $a$ for the observations
using the new spectra. The results are shown in Figure \ref{taubar}.
Increasing $\taueff$ for a given amplitude of mass fluctuations
increases the fluctuations in $\deltaf$,
since it requires us to choose a larger value of $A$ in
equation (2).
As a result, we find a
lower value for $a$.
The $1 \sigma$ error that PRS give on their $\taueff$ measurement corresponds
to $4\%$ at $z=2.5$. This can be translated directly to an error in $a$ of
($+12\%, -9\%$), as shown in Figure \ref{norm}.

In order to combine these two contributions to the 
normalization uncertainty, we first calculate
the likelihood distribution
for the amplitude of mass fluctuations for each of the sources
of error taken individually, assuming that the errors
on $\taueff$ and on the weighted sum $S$ of equation (7) are each Gaussian
distributed. We then convolve the likelihood distributions
and find the total uncertainty, which is $(+20\%, -17\%)$ in the amplitude
of mass fluctuations and  $(+45\%, -31\%)$ in $P(k)$. The combination
of errors is described in more detail in Section 5.1.

The $\taueff$ error is  smaller than the main source of error,
but it is nonetheless important. It would be worth 
investigating the measurement of $\taueff$ in detail,
as a more accurate measurement is critical 
to obtaining more accurate determinations of $P(k)$ using larger samples
of QSO data.
As $\taueff$ in our approach sets the value of the parameter
$A$ in equation (2), it determines how well we
can measure the level of ``bias'' between 
$\tau$ and the mass fluctuations. Measurements of $\taueff$ are also crucial
for constraining the
parameters that are subsumed into $A$, such as $\Omega_{b}$
(see Rauch \etal 1997; Weinberg \etal 1997b).

\begin{figure*}[t]
\centering
\vspace{8.5cm}
\includegraphics{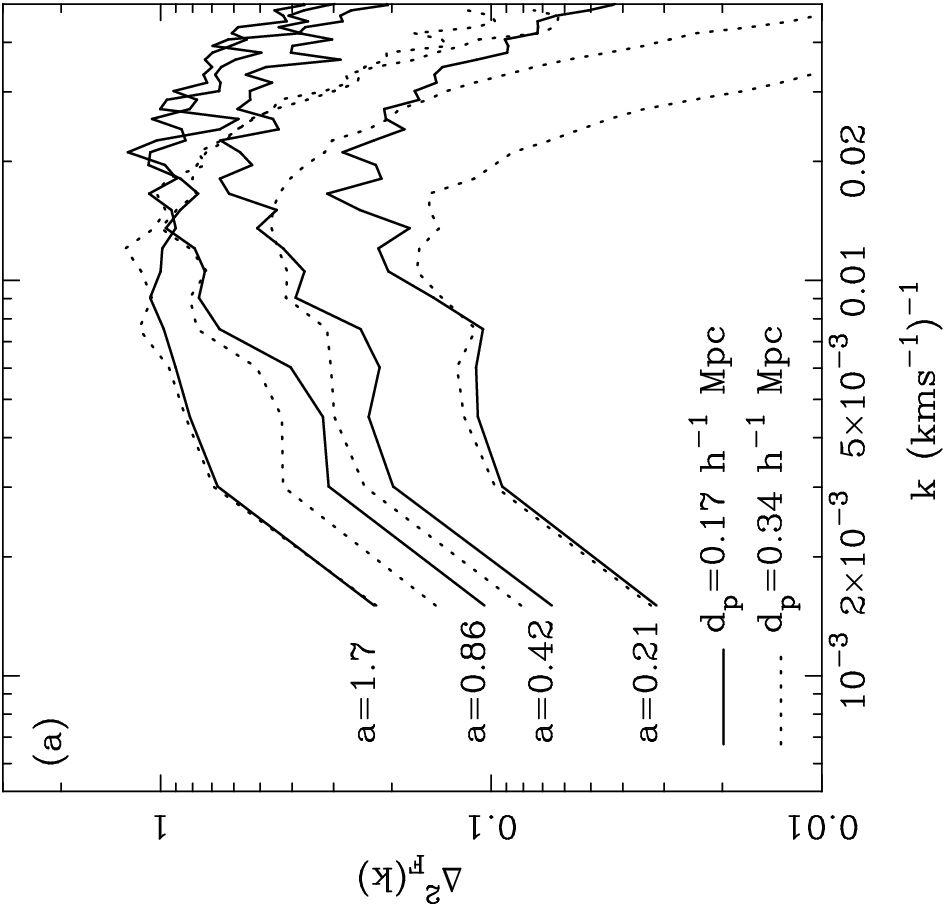}
\includegraphics{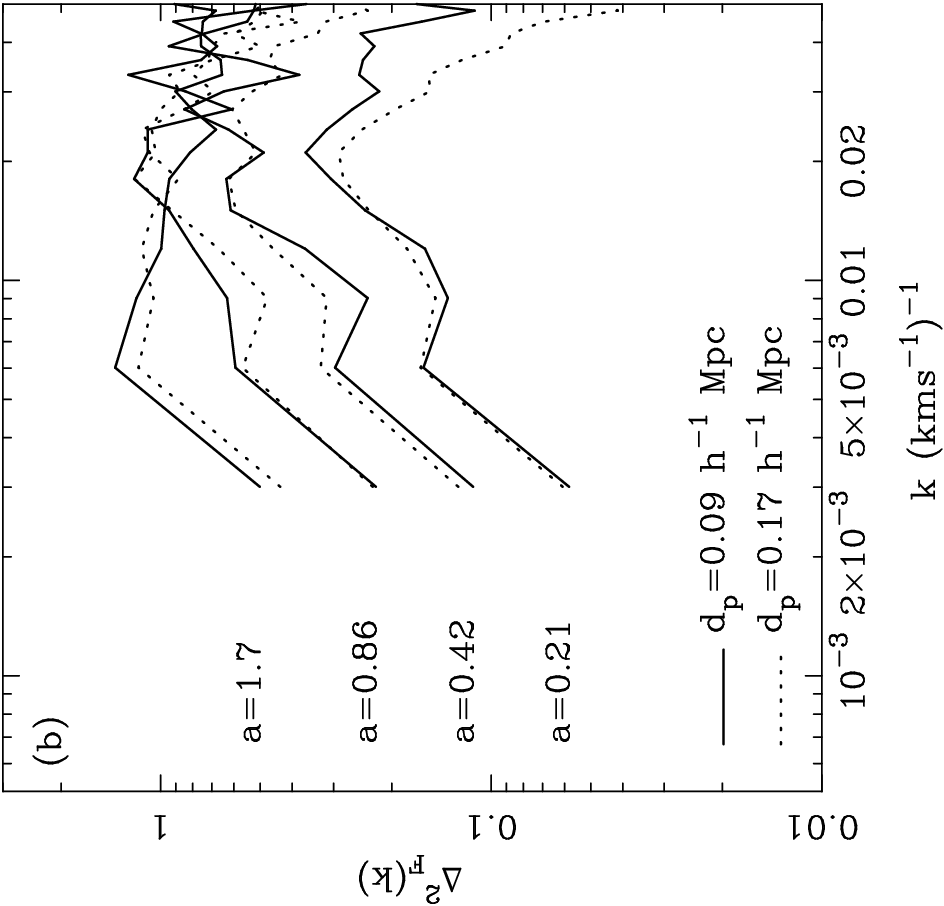}
\caption[junk]{
The effect of varying resolution in the normalizing simulations.
We plot the power spectrum of the flux for a few different output times.
(a) $4170 \kms$ ($22.22 \hmpc$ for $\Omega_{0}=1$) box simulations,
with the same phases, but with different mean interparticle separations
$d_{p}$. The results shown in this panel are
an average of 4 realizations for each resolution.
(b) Results from two single simulations (same phases)
in a $2085 \kms$ ($11.111 \hmpc$ for $\Omega_{0}=1$) box
with different interparticle separations.
\label{resfpk}}
\end{figure*}

Although we do not use $\deltaf$ information on small scales  
in our normalization of $P(k)$, we might also expect the resolution of the
simulations to have some effect on $\deltaf$ on large scales. 
For example, if the normalizing simulations are of insufficient resolution, 
small scale fluctuations that should be
near saturation, or at least away from the linear
part of the curve of growth, might instead be smoothed
out, and therefore contribute more to $\taueff$. The interplay
between $\taueff$ and the amplitude of mass fluctuations described above
would then lead to a systematic offset in the normalization.
To check that our normalizing simulations have sufficient
resolution, we have run some 
simulations with higher resolution and some with lower resolution.
Figure \ref{resfpk}a shows results for the low resolution runs, which use the 
same phases and box size as the original  simulations, but have
only $64^{3}$
particles instead of $128^{3}$. The mean interparticle separation, listed on
the plot legend, is therefore a factor of two larger.  
For two of the plotted output times, this lowering of
resolution has increased $\deltaf$ systematically on large scales.
Normalizing $P(k)$ using these low
resolution simulations would result in a mass fluctuation amplitude
$\sim 20 \%$ lower. 
The effect on $\deltaf$  at smaller scales, $k > 0.015 \invkms$,
is much stronger, since this is the regime 
where the lowered resolution comes directly into play, but these
scales do not enter into our normalization procedure.

Figure \ref{resfpk}b
compares results at our standard resolution to results at higher resolution.
Here we have only run
one realization for each of the two resolutions, with identical phases,
in a box of side length $11.11  
\hmpc$. As the phases are different from the panel (a) runs, 
and the cosmic variance errors are large (the volume
of space simulated is $1/32$ of that in panel [a]), we cannot
compare panels (a) and (b) directly. We can compare the dotted curves, which
have the same resolution as the original simulations, to the solid curves,
which show the effect of increasing
the spatial resolution by a factor of two. This time there is no systematic
offset between the two, so it appears that our  original simulations
have sufficient resolution. The standard normalizing simulations 
have the same mass resolution as the SPH simulations 
analyzed in CWKH (but lower gravitational force resolution),
and each has eight times the volume.
There is a systematic difference between the standard and high
resolution simulations of Figure~\ref{resfpk}b at high $k$,
suggesting that the depression of the small scale $P(k)$
found in the CWKH tests is caused at least in part by
the finite mass resolution of the SPH simulations.

\section{The effect of fluctuations in the ionizing background}

Before examining and discussing our $P(k)$ results in more detail, we
investigate another potential source of systematic error,
clustering in the flux caused by fluctuations in the ionizing background.
If the ionizing background is not uniform, as we have assumed, but instead
exhibits substantial inhomogeneities, then fluctuations in $\tau$ will be 
caused by fluctuations in the spatially varying value of $\Gamma$ in 
equation (2), as well as by fluctuations in the mass density. 
The UV background (UVBG) is produced mainly
by discrete sources, such as QSOs and starburst galaxies.
Whether this discreteness has an important 
effect on our $P(k)$ determination depends on the scale and amplitude
of the clustering induced by the non-uniformity of the UVBG
compared to that produced by intrinsic clustering in the mass. By
claiming that we are able to measure $P(k)$ for the mass, we are effectively 
assuming that the UVBG is uniform on the scales $< 10 \hmpc$ 
that we can access with our current observational data.

Previous work on this issue has examined the effect of UVBG
fluctuations on randomly distributed \lya\ clouds (Zuo 1992; Fardal
\& Shull 1993). In this paper we simulate the fluctuations caused in
uniformly distributed gas, using the FGPA, and also the effect of modulating
observed QSO spectra with additional UVBG 
fluctuations derived from simulations. The case for which we expect there to be
the largest fluctuations is a UVBG entirely generated by QSOs, which
have a very low space density. We will therefore deal with this case first and
in most detail.

Our UVBG simulations are set up in a universe
with $H_{0}=50 \kmsmpc$, $\Omega_{0}=0.2$, and $\Lambda_{0}=0$, 
and a box of size 370 proper Mpc at $z=2.5$ (which corresponds
to $79310\kms$). We populate this box with QSOs, using
luminosities drawn from the
luminosity function of Haardt \& Madau (1996), with a lower cutoff at 
$M_{B}=-23$. We try simulating both Poisson distributed and clustered 
QSO distributions. The clustered QSO positions are chosen by first
generating a Gaussian linear density field in the box using a power 
spectrum appropriate for a $H_{0}=50 \kmsmpc$, $\Omega_{0}=0.2$ CDM model
(taken from Efstathiou, Bond, \& White 1992).
We  select all regions in this field that have a density  above a certain
threshold and populate them randomly with QSOs. As shown by Kaiser
(1984), thresholding produces a distribution
of QSOs that are clustered more strongly than the underlying mass.
We choose the threshold height so that the
scale at which the autocorrelation function of the QSOs is 
unity is $r_0=10\hmpc$ (comoving).

The factor that most strongly influences the level of UVBG fluctuations
is the attenuation length of the QSO flux.
At high redshifts,
the intergalactic medium has a substantial optical depth
to ionizing photons.
Fardal \& Shull (1993) recommend parameterizing the attenuation of 
ionizing 
radiation by intergalactic gas with an attenuation length,
$r_{\rm att}$,  defined so that radiation reaching a  distance
$r$ from a source is attenuated on average
by a factor $e^{-(r/r_{\rm att})}$.
Fardal \& Shull (1993) and Haardt \& Madau (1996) estimate that 
$r_{\rm att} \simeq 100 $ proper Mpc (for $h=0.5$) at $z=2.5$. 
The attenuation length rises rapidly
with increasing redshift, as the Universe becomes more optically thick.
We will try using both $r_{\rm att}=100 $ Mpc and $r_{\rm att}=50 $ Mpc in our 
simulations.  

To generate spectra from our UVBG simulations, we randomly select lines of
sight through the box and calculate the intensity of UV radiation
at each point along them, summing the contributions of
all the QSOs in the box. We assume Euclidean space,
which should be a good approximation
at high redshift, and periodic boundary conditions.
We therefore apply an inverse-square law to the radiation, which is 
additionally attenuated according to the attenuation law described above.
Because of the finite box size, we cut off the flux after it has traveled
one full box side length. The optical depth at each point in the spectra
is calculated according to equation (2),
with $\rho_{b}(x)=1$ (the cosmic mean)
and $\Gamma(x) \propto J(x)$, where $J(x)$ 
is the UV radiation
intensity at point $x$. The value of $A$ is set so that  
$\taueff$ for the spectra is equal to the PRS value of 0.28.

\begin{figure*}[t]
\centering
\vspace{7.7cm}
\includegraphics{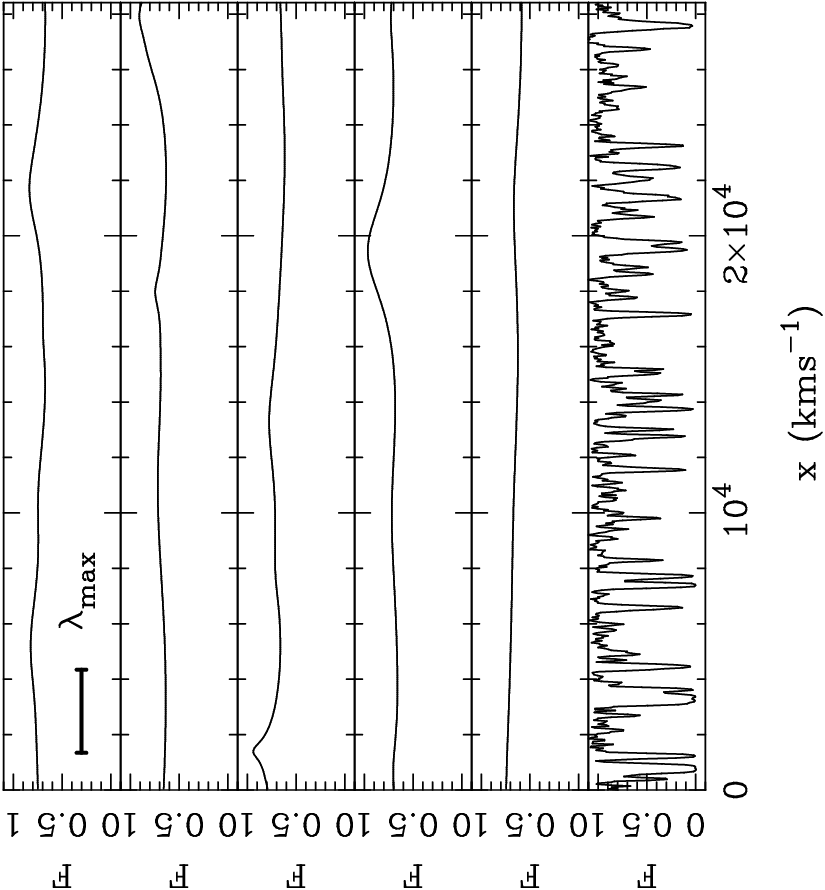}
\caption[junk]{
Top five panels: Lines of sight through a universe with $h=0.5, \Omega_{0}=0.2,
\Lambda=0$, in which a uniform IGM is photoionized by a discrete population of 
QSOs drawn from the Haardt \& Madau (1996) luminosity function. 
The lines represent the transmitted flux assuming the PRS value of
$\taueff$, clustered QSOs, and a 50 Mpc attenuation length, $r_{\rm att}$.
The full simulation occupies a 370 proper Mpc box, 
which is 79310 km/s on a side at $z=2.5$. 
The bottom panel displays a portion of the 
spectrum of Q2206-199. The bar in the top panel is of length $3000 \kms$,
the largest wavelength plotted in the previous $P(k)$ figures.
\label{discrete}}
\end{figure*}

In Figure \ref{discrete}, we show portions of five
sample UVBG simulation spectra (the simulation box
is more than twice the length of the spectra shown), together with a piece
of the spectrum of Q2206-199.
The model spectra represent a universe in which the IGM is uniform
density and absorption fluctuations are caused only by inhomogeneities
of the UVBG. The fluctuations are mild and have a large coherence
scale, very different from the 
observed spectrum shown in the bottom panel.
The bar shown in the top panel is
of length $3000 \kms$, corresponding to the largest wavelength for which 
we have tried to measure $P(k)$. 
The fluctuations caused by UVBG inhomogeneity are small compared to
the observed flux variations on this scale (a point that we will
demonstrate quantitatively below).
It was shown in Section 3.1 that this scale 
is already larger than the minimum scale that is affected by 
continuum fitting. 
Looking at Figure \ref{discrete},
it seems as though UVBG fluctuations  will therefore not
limit our ability to measure $P(k)$ on these scales and below.
We note that the assumptions employed in Figure \ref{discrete}
(strongly clustered QSOs and an attenuation length half the expected value)
are those that tend to maximize the fluctuations.   

\begin{figure*}[b]
\centering
\vspace{7.7cm}
\includegraphics{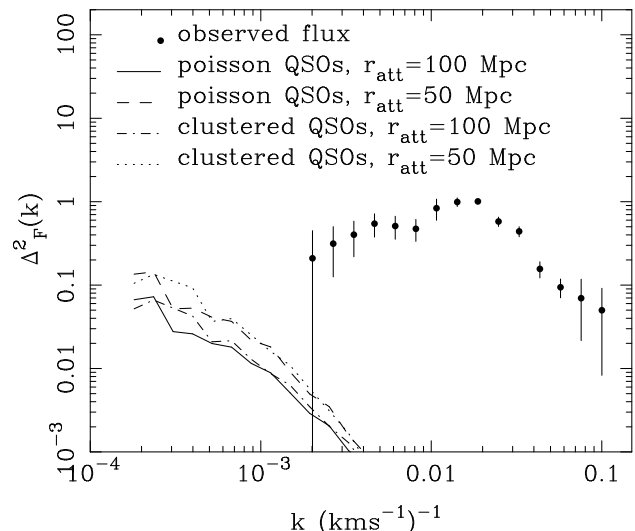}
\caption[junk]{
The 3D flux power spectrum for lines of sight through a uniform density medium
with the PRS value of $\taueff$ and an inhomogeneous UVBG.
We show results for clustered and Poisson distributed QSOs with two 
different values of the attenuation length.
\label{discretepk}}
\end{figure*}

We can examine the effect of UVBG fluctuations
quantitatively by measuring $\deltaf$ for the
UVBG simulation spectra. The results are plotted in Figure \ref{discretepk},
for Poisson
distributed QSOs and clustered QSOs, and for the two different values
of $r_{\rm att}$. Reducing the attenuation length by a factor of two has a
significant effect, raising $\deltaf$ by roughly a factor of two, while
the change induced by clustering the QSOs is barely measurable.
The value of $\deltaf$ in the UVBG simulations is $5 \%$ or less of 
$\deltaf$ in the observations for the largest scale plotted, and $\sim 1\%$ 
for the largest scale reliable enough to use in our analyses,
$k= 2.7 \times 10^{-3} \invkms$. 
On very large scales ($\lambda\sim 100-200 \hmpc$ for
$\Omega=1$), the UVBG fluctuations should become important,
but these scales are beyond the range of the techniques we are using here.  
The results of our analysis agree
with the conclusions reached by Fardal \& Shull (1993), that the UVBG
fluctuations have a relatively small amplitude and a large coherence scale.
We should bear in mind that at higher redshifts, if QSOs are the dominant
source of UVBG radiation, then the fluctuations should increase,
perhaps to a detectable level, because of 
the observed decrease in the space density of QSOs past $z=3$
(Warren \etal 1994)
and because of the decreasing transparency of the IGM.

\begin{figure*}[t]
\centering
\vspace{7.7cm}
\includegraphics{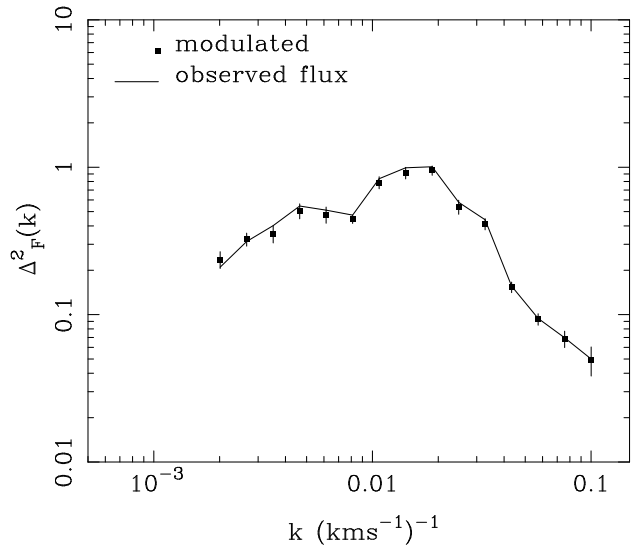}
\caption[junk]{
The 3D power spectrum of the observed QSO spectra modulated by the
inhomogeneous UV background taken from the UVBG
simulations used  in Figure \ref{discrete} 
(clustered QSOs, $r_{\rm att}=50$ Mpc).
The error bars represent the variance measured from modulating the
observations ten times using different realizations of the simulated UVBG.
The lines represent our fiducial result (see previous figures).
\label{modulate}}
\end{figure*}

We can look at the effects of inhomogeneity in the UVBG 
in a different way by modulating
observed QSO spectra with additional fluctuations derived from
our UVBG simulations. This should give us an idea of what occurs when density
fluctuations and UVBG fluctuations 
are taken together. We have done this by taking 
values of $J$ along lines of sight in the UVBG simulations and multiplying
$\tau$ in the observations by $\overline{J}/J(x)$, where $\overline{J}$
is the mean radiation 
intensity. The results are plotted in Figure \ref{modulate},
with error bars
representing the variance of the  results for 10 separate realizations of
the UVBG. There are only very small changes with respect to the 
unmodulated results. The variance between realizations
is apparent, even on small scales. The effects of UVBG fluctuations on small 
scales are probably due to their manifestation  as changes in the 
mean optical depth, which slightly increase the variance in the measured 
$P(k)$ from spectrum to spectrum.

Our UVBG simulations could be extended to include other potential
characteristics of the QSO population. For example, it is possible that QSOs
emit their \lya\ radiation in a highly beamed way, or else that they have 
short lifetimes compared to the light travel time across $r_{\rm att}$.
If either of 
these effects were operating,
the effective space density of QSO sources responsible
for the UVBG should be larger than that 
given by the luminosity function we have 
used. This increased space density
would counterbalance any extra inhomogeneity in the radiation
emitted by the sources themselves.

If the sources of radiation are more numerous than QSOs, we expect the
UVBG to be more homogeneous. We can make a rough estimate of the
size of fluctuations using the model of Fardal \& Shull (1993,
see also  Kovner \& Rees 1989). 
At a certain distance from a source, $r_{p}$, the UV intensity
produced by the source is equal to that produced by the UVBG, so that
 $J_{S}(r_{p})=\overline{J}$. We define the effect of this local UV radiation
to be strong if $r<r_{p}$.
 If all sources have the same luminosity, and a space density $\overline{n}$,
 then  $r_{p}=(4\pi\overline{n}r_{\rm att})^{1/2}$. The volume filling factor,
$f$,
of regions where the effect of local UV radiation is strong is 
$4\pi r_{p}\overline{n}/3$. For small $f$, the variance of UVBG fluctuations
is approximately equal to $f$. If the sources of UVBG radiation are
starburst galaxies with mean separations of 
$5 \hmpc$ comoving, then, using Fardal \&
Shull's value for $r_{\rm att}$ at $z=2.5$, we find $f \sim 10^{-3}$
(if $\Omega_{0}=1$). This is extremely small, showing that the contribution
to the UVBG of galaxies and other sources with the same or greater
 number density will be very smooth. This smoothness will be enhanced by 
re-emission of UVBG radiation from \lya\ forest and Lyman limit systems,
which are even more numerous. This 
smoothing effect
was pointed out by Haardt \& Madau (1996), 
who estimated that recombination radiation
from these systems contributes about $30\%$ of the UVBG at $z=3$.

In this Section we have studied the possible effect of an
inhomogeneous UVBG,
something that has not previously been incorporated into simulations
of the \lya\ forest. 
Other physical processes such as quasar outflows or supernova
shock heating of the IGM on the outskirts of galaxies
could also affect the clustering seen in \lya\ spectra,
but most of these would be confined to high density regions with 
a small volume filling factor.  They would therefore have little
impact on the large scale mass clustering inferred from the \lya\
forest, just as shock heating, collisional ionization, and star formation,
processes that are included in SPH simulations but not in the PM
approximation, have negligible impact on the recovery of $P(k)$
(see CWKH).  A physical effect that could have an impact in low
density regions is inhomogeneous heating of the IGM during
helium reionization (\cite{miralda94}), which could produce spatial
fluctuations in the temperature-density relation if helium reionization
is sufficiently late and sufficiently patchy.
However, if we consider equation~(2) in the limit of weak fluctuations,
we see that the fluctuations in temperature at fixed density would need
to be more than twice as large as the fluctuations in density on
the same spatial scale in order to have an equal effect, since 
$\tau \propto \rho_b^{1.6} T_0^{-0.7}$.

In the long run, rather than trying to investigate all possible sources of
spurious clustering, we should look for support for the gravitational
instability interpretation of \lya\ forest fluctuations  
in the observational data themselves.
Already many aspects of the observational \lya\ forest data  can be
reproduced and explained by the scenario
(see, e.g., Bi \& Davidsen 1997; Rauch \etal 1997).  One specific test of 
our approach is the measurement of the evolution of $P(k)$ with redshift.
The $P(k)$ we measure should change in the way predicted by linear theory,
keeping the same shape and increasing in amplitude in a way
that  (in detail) depends on $\Omega_{0}$ and $\Lambda_{0}$. 
It seems very unlikely that any non-gravitational processes could precisely
mimic this behavior, so if linear growth were seen in the data it would
provide strong evidence for the validity of the $P(k)$ measurement. 
Although the observational sample we are using in this paper is too small
to carry out this test unambiguously, we are at least 
able to split the sample into two redshift halves and see if
there are any gross deviations from linear growth. We will do this 
below.

\section{Results}

\subsection{Tabulation of $P(k)$ and a power law fit}
In Table 1, we give the values of $P(k)$ and their $1 \sigma$ errors for the 
seven points where we believe that our measurement is representative of
its primordial value. The $1 \sigma$ errors were calculated in Section 3.1,
from the scatter between results for 10 subsamples of the data. They 
primarily represent uncertainties in the {\it shape} 
of $P(k)$, as there is a separate normalization
uncertainty that applies to all points equally. This normalization
uncertainty was estimated 
in Section 3.2, again from the scatter in results 
between 10 subsamples of the data. The covariance matrix of the data values
(calculated using the 10 subsamples) has some non-negligible off-diagonal
terms, which quantitative evaluation of models should take into account.

\begin{table*}[h]
\centering
\caption[junk]{\label{pktab} The linear $P(k)$ at $z=2.5$. 
We give the wavenumber $k$, $P(k)$, and the $1 \sigma$ error
on $P(k)$. An additional error should also be assigned to the normalization
of all points, which is $+40\%, -29\%$ in $P(k)$ ($1 \sigma$). }
\vspace{1cm}
\begin{tabular}{ccc}
\hline &&\\
$k$ $\invkms$ & $P(k)$ $(\kms)^{-3}$ & $\sigma$[$P(k)$]  \\
\hline &&\\
$2.66 \times 10^{-3}$ & $3.6 \times 10^{8}$  & $1.9  \times 10^{8}$  \\
$3.52 \times 10^{-3}$ & $1.7 \times 10^{8}$  & $8.9  \times 10^{7}$  \\
$4.65 \times 10^{-3}$ & $9.4 \times 10^{7}$  & $2.8  \times 10^{7}$  \\
$6.14 \times 10^{-3}$ & $3.4 \times 10^{7}$  & $1.0  \times 10^{7}$  \\
$8.12 \times 10^{-3}$ & $1.7 \times 10^{7}$  & $5.1  \times 10^{6}$  \\
$1.07 \times 10^{-2}$ & $1.3 \times 10^{7}$  & $3.8  \times 10^{6}$  \\
$1.42 \times 10^{-2}$ & $6.2 \times 10^{6}$  & $1.1  \times 10^{6}$  \\
\hline &&\\
\end{tabular}
\end{table*}

As the estimated errors on our points are fairly large, and we cover a limited
range in $k$, the information in our $P(k)$ measurement can be
effectively summarized by the
amplitude and slope of a power law fit to the data points.
When determining the parameters of this fit, we can include the
effect of covariances between data points, which are not given in the Table
above. To eliminate as much as possible the covariance between the
fit parameters themselves, we have chosen to describe the amplitude
of the fit by the value of $P(k)$ at a pivot
wavenumber $k_{p}$ near the center of the data range.
The form we fit is therefore
\begin{equation}
P(k)=P_p\left(\frac{k}{k_{p}}\right)^{n}.
\end{equation}
We perform a $\chi^{2}$ fit to the seven data points, including the
full covariance matrix for the points [as in 
equation (6), though here it is the Gaussianized flux $P(k)$
rather than $\deltaf$ that enters]. We try several values for the
pivot wavenumber and choose $k_{p}=0.008 \invkms$, the value
for which the covariance between $P_{p}$ and $n$ is minimized.
In evaluating the covariance matrix of the $P(k)$ data points
from the ten subsamples of the full data set, we find that the fluctuations
of neighboring data points are usually anticorrelated, probably
because of the differentiation involved in going from the 1D power
spectrum to the 3D power spectrum (equation [3]).
As a consequence, the statistical error on the power law
slope $n$ is smaller than it would be if we ignored the covariances
in our $\chi^2$ evaluation.
Because the anticorrelated structure of the covariance
matrix significantly influences the error estimate and the
estimate of the covariance matrix from the data subsamples is
itself noisy, we regard 
our estimate of the error on $n$ as itself significantly uncertain.
If we ignored covariance terms when fitting $n$, we would get error bars
$\sim45 \%$ larger ($n=-2.30\pm0.26$) than
those reported below based on using the full covariance matrix.

\begin{figure*}[t]
\centering
\vspace{7.7cm}
\includegraphics{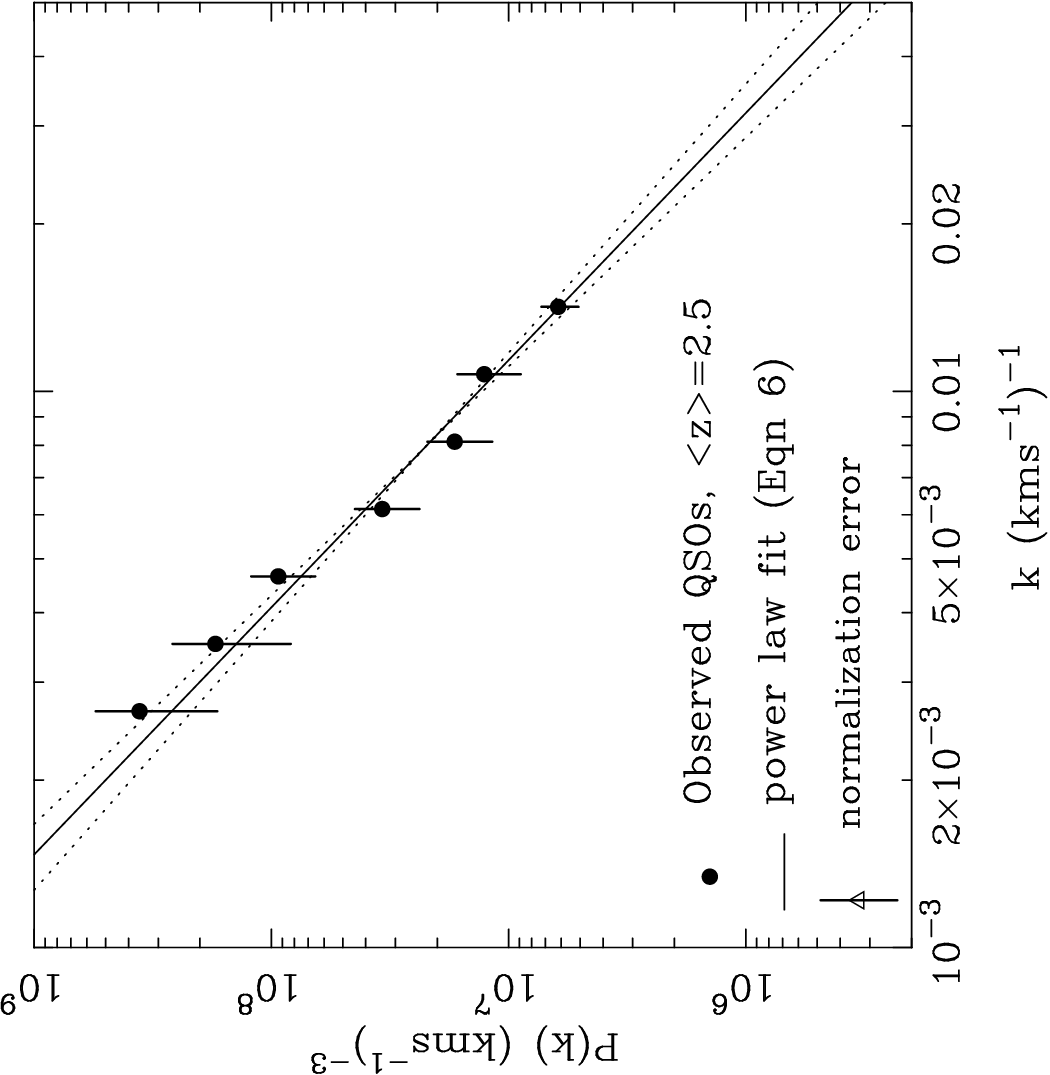}
\caption[junk]{
The fiducial $P(k)$ result at $z=2.5$. Points plotted are those from Table 1.
Also shown is the power law fit of equation (8), together with dotted lines 
showing the $\pm 1 \sigma$ uncertainty in the slope for fixed $P(k_p)$.
The error bar at lower left shows the $1\sigma$ normalization uncertainty.
At the $1\sigma$ level, all of the data points can be shifted
coherently up or down by this amount.
\label{powerlaw}}
\end{figure*}

\begin{figure*}[b]
\centering
\vspace{8.0cm}
\includegraphics{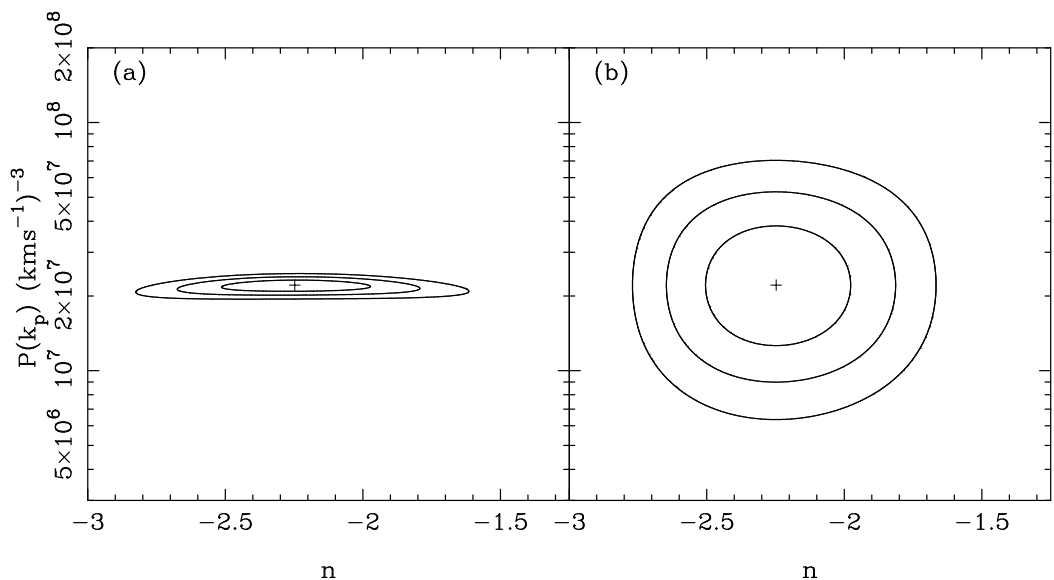}
\caption[junk]{
(a) Contours of constant $\Delta\chi^{2}$ resulting from fitting a
power law (eq.~[8])
to the $P(k)$ data for the fiducial sample.
The amplitude of $P(k)$ at the pivot wavenumber, $k_{p}=0.008 \invkms$,
is shown on the $y$-axis, and the logarithmic slope, $n$, on the $x$-axis.
The best fitting values are marked by a cross, and the contours enclose 
$68\%$, $95 \%$ and $99.7\%$ of the joint probability.
(b) Contours of 68\%, 95\%, and 99.7\% joint probability
($\Delta\chi^2$=2.30, 6.17, 11.80)
after including the overall normalization uncertainty
(see text and Figure~\ref{chi2_1d}).
\label{chi2_2d}}
\end{figure*}

Figure \ref{powerlaw} shows the best fit power law, together with the
points from Table 1.
Figure \ref{chi2_2d}a shows contours
of constant $\Delta\chi^{2}$ for the fit, where
$\Delta\chi^{2}=\chi^{2}(P_{p},n)-\chi^{2}(P_{p\min},n_{\min})$
and $(P_{p\min},n_{\min})$ are the values of the fit parameters for which
the $\chi^{2}$ is a minimum. The 1,2 and $3\sigma$
contours  of joint confidence in 
the fit parameters taken together are shown,
corresponding to $\Delta \chi^2 = 2.30$, 6.17, 11.80.
The value of $\chi^{2}$ at the minimum is 4.0. A value greater than this would
be expected to occur $55 \%$ of the time given that we have five degrees of 
freedom (seven data points minus two free parameters).
We can see that for  the pivot wavenumber we have chosen,
the errors on the slope and amplitude of
the power law fit are effectively independent.

\begin{figure*}[t]
\centering
\vspace{8.0cm}
\includegraphics{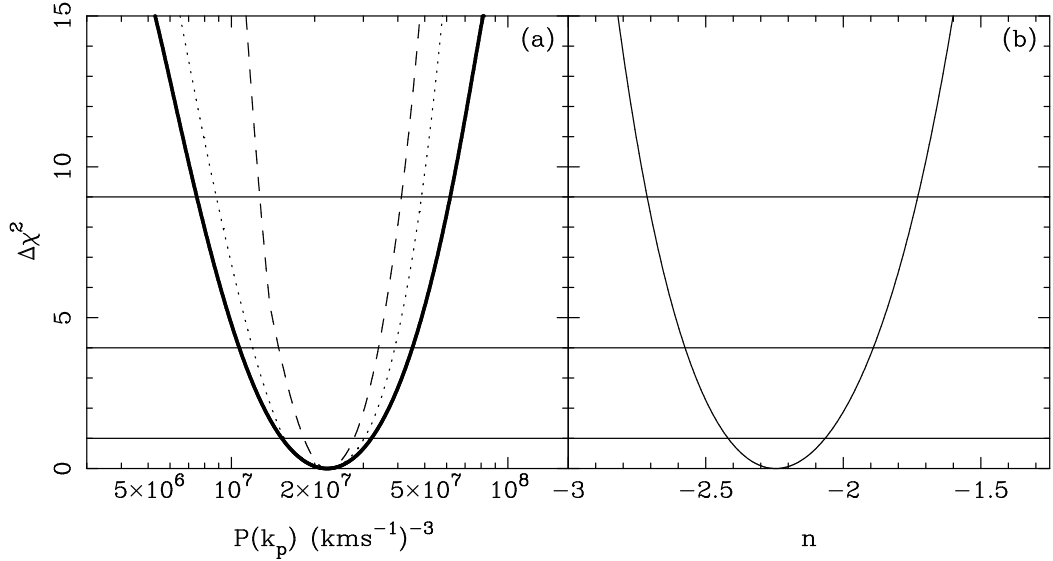}
\caption[junk]{
One-dimensional $\Delta\chi^{2}$ distributions
resulting from a power law fit to $P(k)$ for the fiducial sample.
(a) The uncertainty in the amplitude $P(k_p)$.
The dotted line shows the uncertainty from the normalization
procedure, corresponding to the error bar in the lower left
corner of Figure~\ref{powerlaw}.  The dashed line shows the
uncertainty from the error in $\taueff$ (Figure~\ref{taubar}).
The heavy solid line shows the combined uncertainty, obtained
by convolving the two likelihood distributions (assumed to be Gaussian).
(b) The uncertainty in the power law slope $n$.
\label{chi2_1d}}
\end{figure*}

The uncertainty in $P_p$ comes not from the uncertainty in fitting
a power law to the $P(k)$ data points but from the normalization
uncertainty detailed in \S 3.2, which affects the level of all the
data points simultaneously.  The amplitude of the mass power spectrum
is fixed by requiring that spectra from the normalizing simulations
reproduce the observed value of the amplitude diagnostic $S$
(equation [7]), and the uncertainty is determined from the uncertainty
in $S$ estimated from the scatter among subsamples.
There is an additional contribution to the normalization uncertainty
from the uncertainty in $\taueff$, as illustrated in 
Figure~\ref{taubar}.  To combine the two sources of error, we construct
$\Delta\chi^2$ distributions for each (shown by the dotted and
dashed lines in Figure~\ref{chi2_1d}a), assuming that the
errors on $S$ and $\taueff$ are Gaussian distributed.
We then convolve the two corresponding likelihood distributions,
${\cal L}/{\cal L}_{\rm max}=e^{-\Delta\chi^{2}/2}$
where ${\cal L}_{\rm max}$ is the maximum likelihood, and convert
the convolved likelihood distribution into a combined $\Delta\chi^2$
curve, shown by the thick line in Figure~\ref{chi2_1d}a.
The intersection of this curve with the horizontal lines at
$\Delta\chi^2=1,$ 4, 9 gives the $1\sigma$, $2\sigma$, and $3\sigma$
errors on $P_p$.  The uncertainty coming from the normalization
procedure dominates the overall uncertainty in $P_p$, but the
$\taueff$ uncertainty makes a significant contribution and could
easily come to dominate in the analysis of a larger data set.
We do not include the uncertainty in the power law fit amplitude
as a separate source of error because it has already been counted in
the normalization error --- the uncertainty in the overall level
of the data points is the reason for uncertainty in the amplitude
diagnostic $S$.

Our final results for the power law parameters and their $1\sigma$ errors are
$P_{p}=2.21^{+1.00}_{-0.68}\times 10^{7} (\kms)^{-3}$
and $n=-2.25^{+0.18}_{-0.18}$.
The $\Delta\chi^{2}$ distribution for 
$n$ is shown in  Figure \ref{chi2_1d}b.
As discussed in \S 3.2, if we had used a maximum likelihood fit
to normalize $P(k)$ instead of our more conservative (and, we think,
more robust) method based on the diagnostic $S$, we would have
obtained a value of $P_p$ 30\% (0.7$\sigma$) higher and statistical
uncertainties in $P_p$ smaller by $\sim 30\%$ (after including the 
$\taueff$ error).

Because the errors on $P_{p}$ and $n$ are independent,
we can combine their one-dimensional $\Delta\chi^{2}$ distributions 
into a two-dimensional plot by simply adding the 
the values of $\Delta\chi^{2}$
(equivalent to multiplying the likelihoods). The $\Delta\chi^{2}$ contours 
corresponding to 68\%, 95\%, and 99.7\% confidence intervals for
$\chi^2$ distributions with two degrees of freedom
are shown in Figure \ref{chi2_2d}b.

Since the off-diagonal terms in the covariance matrix of the $P(k)$
data points are significant, and the values and uncertainties of
$P_p$ and $n$ summarize the results effectively, we recommend
use of the power law fit parameters rather than the tabulated $P(k)$
when evaluating models.
A quantity whose physical meaning is more intuitively obvious than
$P_{p}$ is  $\Delta^{2}_{\rho}(k_{p})$,
the contribution to the variance of density fluctuations from a 
logarithmic interval in $k$, given by
\begin{equation}
\Delta^{2}_{\rho}(k_{p})=\frac{1}{2\pi^{2}} k_p^{3} P_{p}
\end{equation}
(see, e.g., \cite{peacock94}).
Our results in terms of this quantity are
$\Delta^{2}_{\rho}(k_{p})=0.57^{+0.26}_{-0.18}$ (1 $\sigma$ errors).

\subsection{Redshift evolution of $P(k)$}
\begin{figure*}[t]
\centering
\vspace{11.3cm}
\includegraphics{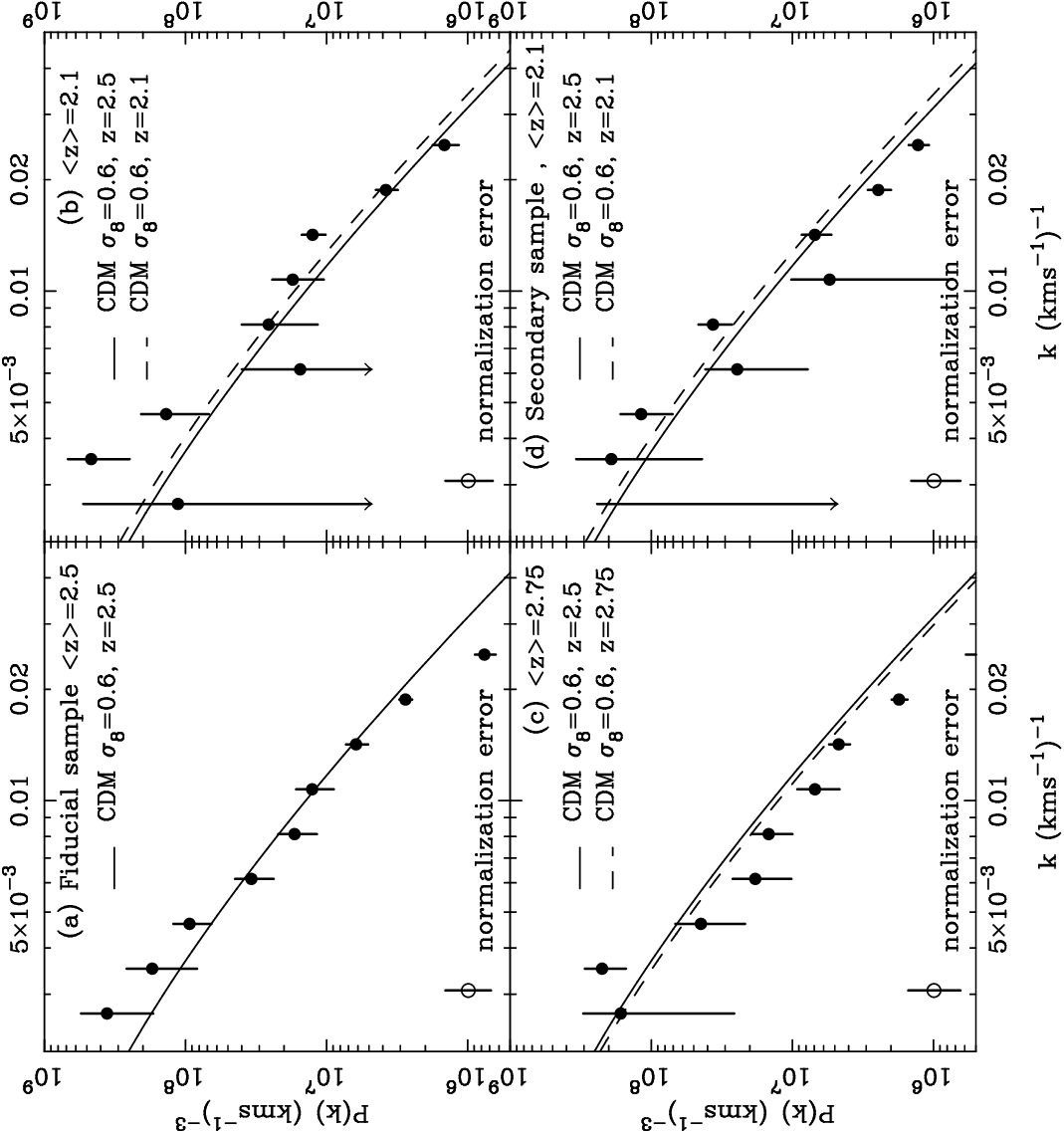}
\caption[junk]{
Redshift evolution of $P(k)$.
Panel (a) shows $P(k)$ for our fiducial sample at $z=2.5$, together with a 
$\sigma_{8}=0.6, \Omega_{0}=1$ CDM model at the same redshift, for reference.
In panels (b) and (c) we have split the whole sample 
(all the spectra plotted in
Fig.~\ref{spectra}) into two halves, which have the mean redshifts given in the 
plot labels. The $\sigma_{8}=0.6$ CDM model is again shown, this time 
at $z=2.5$ and at the appropriate $z$ for the subsample. Panel (d) shows
results from a wholly independent sample of 9 QSO spectra in a 40 arcmin
AAT field.
The mean redshift of this sample is about the same 
as the low-$z$ half of the main sample, and the effective number of 
QSOs is also about the same as in the low-$z$ half of the main sample.
The different samples are described in more detail in Section 2.1.
\label{zevol}}
\end{figure*}

We now test the effect of redshift evolution using the 
high- and low-$z$ halves of our data (see Section 2.1 for a description of the
samples).  When the linear $P(k)$ evolves with redshift, it is subject to two
main effects. First, there is the change in the amplitude of $P(k)$ 
owing to linear growth, with $P(k)$ increasing in proportion to the linear 
growth factor squared as $z$ decreases.
The growth factor is proportional to $a(t)$ in an
Einstein-de Sitter model, and in other models its evolution 
depends on the values of $\Omega_0$ and $\Lambda_0$
(see, e.g., Peebles 1980). 
In a plot of $P(k)$ against $k$, such as Figure \ref{zevol}, 
the linear growth of $P(k)$ would affect only the $y$-axis.
However, because our observed units are velocities rather than
comoving distances, the evolution of the Hubble parameter, $H(z)$, 
shifts $P(k)$ along both the $x$- and $y$-axes.
The $z$ dependence of $H(z)$ is determined by $\Omega_{0}$
and $\Lambda_{0}$ through the Friedmann equation, which can
be rearranged to yield
\begin{equation}
H(z)=H_{0}\left[\Omega_0(1+z)^{3}+(1-\Omega_{0}-
\Lambda_{0})(1+z)^{2}+\Lambda_{0}\right]^{1/2}.
\end{equation}

In decelerating universes the $P(k)$
curve shifts to the right as $z$ decreases because
a given scale in units of comoving $\hmpc$
corresponds to a smaller scale in $\kms$. 
Because $P(k)$ is also in velocity units,
the change of scale also shifts the $P(k)$ curve downwards. 
These changes in units
partially cancel the linear growth of $P(k)$, so
the overall measured $z$-evolution of $P(k)$ is expected to be
rather weak. We will therefore need a large observational data sample
and a long $z$ baseline to discriminate between
models with different values of $\Omega_{0}$ and $\Lambda_{0}$.
With our current data we will restrict ourselves to the less ambitious
goal of testing whether measurements of
$P(k)$ from the two different redshift subsamples 
are consistent with linear growth. The relatively small sizes of our samples
do not allow us to measure the growth rate, but 
we could potentially detect the consequences of a non-gravitational
process that alters the clustering in the \lya\ forest. For example,
if large scale inhomogeneous reheating of the intergalactic
medium occurred during the redshift interval covered by our samples
($z \sim 3.2 \rightarrow 1.6$), and this reheating was important enough to
change \lya\ clustering, then  we would not expect
our  estimates of $P(k)$ for the high-$z$ and low-$z$ samples
to be consistent with linear growth.

In Figure \ref{zevol}a we show the $P(k)$ results for our fiducial ($z=2.5$)
sample. In this Figure, and in those in the
rest of this section, we do not plot the largest scale point
displayed in Figures~\ref{contfitpk}--\ref{norm} because it
was shown in Section 3.1 to be sensitive to continuum fitting uncertainties.
Figure~\ref{zevol}a also
shows the $z=2.5$ linear $P(k)$ of a spatially flat CDM model 
with $\Omega_{0}=1$, $h=0.5$, and $\sigma_{8}=0.6$, where
$\sigma_{8}$ is
the amplitude of density fluctuations in $8\hmpc$ spheres linearly
extrapolated
to $z=0$. The model will be described in more detail in the next section.
At the moment, it serves as a reference curve to which we can compare the
results from the different $z$ subsamples of the data.  

We evaluate $P(k)$ for the high- and low-$z$ halves of the data, subsamples
(3) and (4) of Section 2.1, and for the secondary, fiber field sample,
also described in Section
2.1. We use the same normalizing simulations that were used for the fiducial 
sample, since the  $P(k)$ shape for the subsamples
appears in Figure \ref{zevol} to be consistent with being a noisy version
of the $P(k)$ shape for the  fiducial sample. 
Before normalizing, we must take into account 
that there might be an amplitude offset between the Gaussianized $P(k)$
measured from the different redshift subsamples and the Gaussianized
$P(k)$ from the fiducial $z=2.5$
sample used to set up the normalizing simulations.
We  measure this amplitude offset using equation (7), except that
we replace $\deltaf$ with $P(k)$. This tells us an additional 
factor we must use in our
normalization of $P(k)$ from the different redshift subsamples,
which in all cases turns out to be less than $10 \%$.
When measuring the amplitude offset, we must make sure that 
we are comparing $P(k)$  values on the same scales.
This entails a small rescaling
of the length scales for the different subsamples, where we rescale $k$ 
to the value it would have in $\invkms$ at $z=2.5$ using equation (10). 
We assume $\Omega_{0}=1$ to do this, but our
results are not sensitive to this choice. 
Having done this, and having found the amplitude offset, 
we then carry out the normalizations using $\deltaf$ and 
equation (7), as was done in Section 3.2.

Results for the low and high redshift subsamples are shown in 
Figures~\ref{zevol}b and~\ref{zevol}c, and
results from the secondary, fiber field sample 
are shown in Figure \ref{zevol}d. 
The mean redshift is $\langle z \rangle=2.1$ 
for both the low-$z$ subsample and the secondary sample,
and $\langle z \rangle=2.75$ for the  high-$z$ subsample.
In every panel, the solid curve shows the linear $P(k)$ of
the CDM model at $z=2.5$, while the dashed curves in panels (b)-(d)
show the CDM $P(k)$ at the mean redshift of the sample in question.
The effect of linear evolution is subtle because of the modest
redshift range and the cancellation effects already mentioned.
The linear growth in this model assumes $\Omega_0=1$, but results
would be similar for other cosmological parameters because $\Omega$
approaches one at high redshift in all models.

The $P(k)$ shape for the fiducial sample is consistent with, or 
perhaps slightly steeper than, the $P(k)$ predicted by the CDM model.
The $P(k)$ shapes for the other samples are also consistent with
the model, and hence with being noisier realizations of the $P(k)$
shape measured from the fiducial sample.  The normalization 
uncertainties for the subsamples (shown by the error bars in the 
bottom left of each panel) are significantly larger than the
small $P(k)$ shifts predicted by linear evolution, so we cannot
achieve a positive detection of linear growth with this data set.
However, we do find that the results for the high-$z$ and low-$z$
subsamples are {\it consistent} with linear growth --- in particular, 
there is no significant change in the measured shape of $P(k)$
between $z=2.75$ and $z=2.1$.  It is especially reassuring to see
that the secondary sample, which is wholly independent of our
main sample, gives perfectly consistent results.

We can quantify this consistency by 
carrying out power law fits to the power spectra derived from the 
subsamples, using the procedure described in Section 5.1.
We again use equation (8), with the same pivot wavenumber, $k_{p}=0.008
\invkms$. The value of $\chi^{2}$ per degree of freedom for the best fit power
laws in these cases varies between 2 and 4, which should
only occur $7\%$ and $0.02\%$ of the time, respectively. 
The high $\chi^2$ values probably indicate that
our method of scaling errors to small subsamples from the main sample 
gives errors that are somewhat too small. In the future, with larger data sets,
it will be possible to derive the errors directly from the different redshift
subsamples. Here, to the extent that it is possible to compare results,
we find that the fiducial sample fit parameters fall within or near
the formal 2 $\sigma$
confidence contours for the subsample results, assuming linear evolution
(and $\Omega_{0}=1$). If we consider each parameter 
individually, we find for the $\langle z \rangle=2.1$ subsample,
$\Delta^{2}_{\rho}(k_{p})=0.61^{+0.28}_{-0.23}$ 
and $n=-1.10 \pm 0.55$, and for the secondary sample (also at
$\langle z \rangle=2.1$), $\Delta^{2}_{\rho}(k_{p})=0.47^{+0.22}_{-0.18}$
and $n=-2.81 \pm 0.24$ (all errors are $1\sigma$).
If $\Omega_{0}=1$ we would expect 
$\Delta^{2}_{\rho}(k_{p})=0.70$ for both of these samples,
based on scaling the result for the fiducial sample.
For the $\langle z \rangle=2.75$ subsample, we find 
$\Delta^{2}_{\rho}(k_{p})=0.35^{+0.19}_{-0.15}$ and $n=-2.90\pm0.26$,
where $\Delta^{2}_{\rho}(k_{p})=0.51$ is expected for  $\Omega_{0}=1$.
Because the subsamples are not independent of the fiducial sample, this
comparison is not completely rigorous. We still expect, however,
that any deviation 
from linear growth would have to be fairly small in order to escape detection.
The fact that logarithmic slopes of different subsamples differ
by up to $2\sigma$ suggests that our $\chi^2$ procedure may underestimate
the true uncertainty in $n$.  The agreement of 
$\Delta_\rho^2(k_p)$ values at the $1\sigma$ level suggests
that our estimate of the normalization uncertainty is reasonably accurate.

\subsection{Comparison with theory}
We have already shown in Figure \ref{zevol} the linear theory $P(k)$ for
a CDM model that has roughly the correct shape and
amplitude to match our observed $P(k)$.
In Phillips \etal (1998), we conduct detailed comparisons
of our $P(k)$ results to the predictions of COBE-normalized CDM models,
and we discuss how \lya\ $P(k)$ measurements may be used to
break degeneracies between cosmological parameters that are
left by other measurements (e.g., Efstathiou \& Bond 1998).
In Weinberg et al.\ (1998a) we combine our results with
constraints from the mass function of rich galaxy clusters to
estimate the value of $\Omega_0$.
In this paper, our main emphasis is on the presentation and testing of
our observational results, so we limit our theoretical discussion
to an illustrative and qualitative comparison between our measured
$P(k)$ and the predictions of a few CDM models.

\begin{figure*}[t]
\centering
\vspace{9.4cm}
\includegraphics{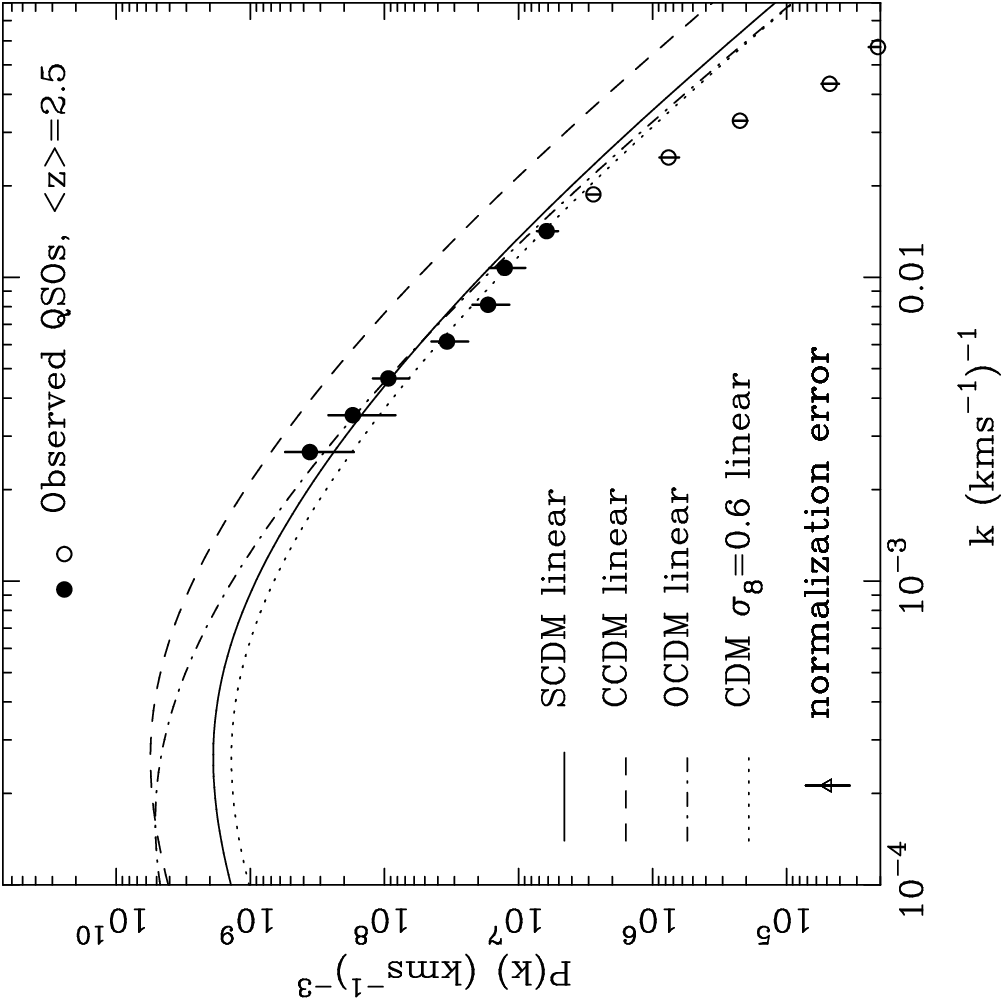}
\caption[junk]{
The normalized $P(k)$ from the fiducial observational sample
compared to 4 different CDM models (see text),
all at $z=2.5$.
Filled circles are plotted on scales where the numerical experiments of
CWKH show that the linear theory $P(k)$ is correctly recovered.
Open circles represent the results on smaller scales.
At the $1\sigma$ level, all points can be shifted up or down coherently
by the normalization error shown in the lower left.
\label{models}}
\end{figure*}

Three of the linear power spectra we compare to
are those used in the hydrodynamic simulations for which we tested
$P(k)$ recovery in CWKH.  All these models have an inflationary power spectrum
with $n = 1$.
The first is SCDM, ``standard'' CDM,  a model  with $\Omega=1$,
$h=0.5$, $\Omega_{b}=0.05$, and $\sigma_{8}=0.7$.
This value of $\sigma_{8}$ is roughly 
consistent with (but somewhat higher than)
that advocated by White, Efstathiou, \& Frenk (1993)
to match the observed masses of rich galaxy clusters,
Our second model is identical to the first
except that $\sigma_{8}=1.2$. This 
higher amplitude is consistent with the 4-year
COBE data (Bennett et al.\ 1996), and we therefore label the model CCDM.
The third model, OCDM, assumes an open universe with 
$\Omega_{0}=0.4$, $h=0.65$, and $\Omega_{b}=0.03$. 
This model is also COBE-normalized, with $\sigma_{8}=0.75$ 
(Ratra et al.\ 1997).
A $\Lambda$CDM model with a modest ``tilt'' of the primeval
power spectrum ($n_p \approx 0.9$) would yield a similar prediction.

The linear power spectra of
these models are plotted in Figure \ref{models}, together
with that of the $\sigma_{8}=0.6$ CDM model already shown in Figure \ref{zevol}.
The measured $P(k)$ is somewhat steeper than that of the SCDM model
and even the OCDM model: the points with $k>4 \times 10^{-3} (\kms)^{-1}$
all lie below the model curves, and the points with 
$k<4 \times 10^{-3} (\kms)^{-1}$ all lie on or above the curves.
However, given the current statistical uncertainties, the
difference in slope is at most suggestive.
Perhaps more impressive is the fact that the linear mass power
spectrum, which has never previously been measured on these scales,
has approximately the slope predicted by the physical model of
inflationary fluctuations in a CDM-dominated universe.
(Studies of galaxy clustering at $z=0$ probe the non-linear rather
than the linear power
spectrum on these scales, and the shape of the galaxy and mass
power spectra could be different because of scale-dependent bias
in the non-linear regime.)

The amplitude of the measured $P(k)$ is somewhat lower than that of
the SCDM and OCDM models, though it is consistent with these
models within the $1\sigma$ normalization error (shown in the lower left).
The $\Omega_0=1$, $h=0.5$, $\sigma_8=0.6$ model appears to have about
the right amplitude.  Since the rms mass fluctuation amplitude,
$\sigma_\rho \propto \sqrt{P(k)}$, is a factor of two larger in the CCDM model,
and the uncertainty in the measured amplitude is only 18\%
(Section 3.2), our results rule out the CCDM model at the $\sim 5\sigma$
level.  Of course this model is already known not to be viable
because it predicts excessively massive galaxy clusters at $z=0$
(e.g., White et al.\ 1993), but that failure 
reflects a combination of the high $P(k)$ amplitude and the high
mass density ($\Omega_0=1$), both of which influence cluster masses.
The present test, based on
independent data at a different redshift, 
shows that the amplitude of mass fluctuations in the CCDM model
is too high regardless of the value of $\Omega_0$.

\subsection{Comparison with observations of galaxy clustering}

The success of recent searches for Lyman Break Galaxies (LBGs,
see \cite{pettini98} for a recent review)
has opened a new window on structure in the high redshift universe:
the clustering of star-forming galaxies at $z \sim 3$
(\cite{steidel98}; \cite{giavalisco98}; \cite{adelberger98}).
The mean redshifts of the LBG samples are close to the mean
redshift of our \lya\ forest data.  We can therefore compare
our measurement of mass clustering to the measurements of galaxy
clustering and obtain a
direct measurement of the bias between galaxy and mass fluctuations
at high redshift.

\begin{figure*}[t]
\centering
\vspace{8.2cm}
\includegraphics{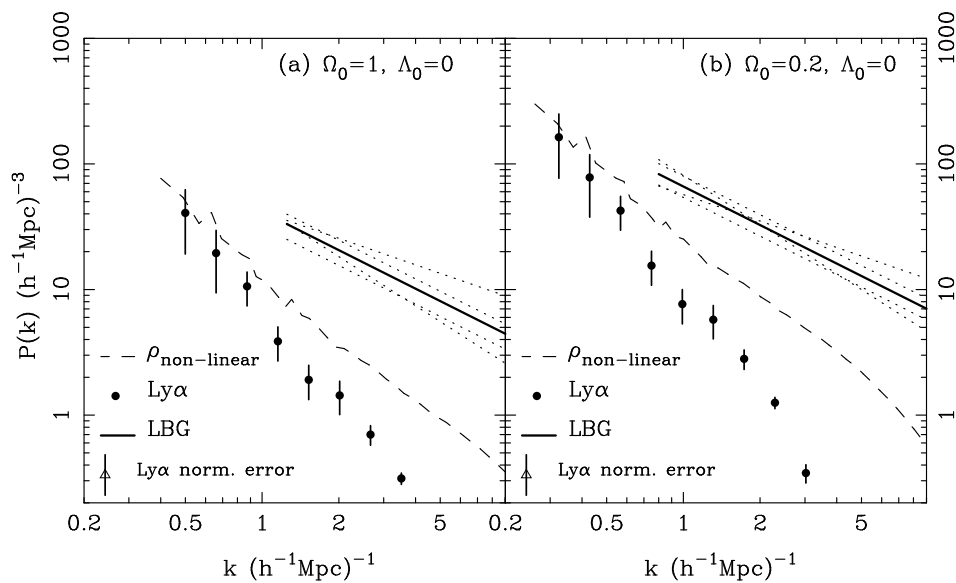}
\caption[junk]{
The normalized $P(k)$ compared to Lyman Break Galaxy clustering
(from Giavalisco \etal 1998), at $z=3.04$.
The solid line shows  the LBG power law fit, and is only plotted  for
scales $k > 2\pi/r_{\rm max}$, where $r_{\rm max}$ 
is the largest pair separation
used in the computation of LBG clustering.
The dotted lines show the effect of varying the LBG amplitude by $\pm 1 \sigma$
whilst keeping the slope fixed and also of varying the slope
by $\pm 1 \sigma$ whilst keeping the amplitude fixed.
Points show the linear mass $P(k)$ derived from the \lya\ forest
data, scaled to $z=3.04$ assuming linear growth.
Dashed lines show the non-linear $P(k)$ measured using a 3D FFT
from the mass distribution in
the normalizing simulations. This non-linear mass $P(k)$ was obtained
by interpolating between results from the two
outputs closest to $z=3.04$. 
\label{lbg}}
\end{figure*}

Giavalisco \etal (1998) have measured the angular clustering of a sample of
871 galaxies in a narrow redshift range centered on $z=3.04$. 
By inverting
the angular clustering using Limber's equation and the estimated
redshift distribution, they obtain an estimate of the real space correlation
function, $\xi(r)$. 
Fitting their results to the power law form exhibited by low-$z$ galaxies,
$\xi(r)=(r/r_{0})^{-\gamma}$, they find $\gamma=1.80^{+0.25}_{-0.19}$ and
$r_{0}=2.1^{+0.4}_{-0.3} \hmpc$ for $\Omega_{0}=1$ or 
$r_{0}=3.3^{+0.7}_{-0.6} \hmpc$ for
$\Omega_{0}=0.2$, $\Lambda_{0}=0$.
To compare these results with $P(k)$ measurements, we have converted
the power law fit to $\xi(r)$ into a power law in $P(k)$, 
$P(k)=Ck^{n}$, using the fact that, for $-2 < n < 0$,
\begin{equation}
\xi(r)=\sqrt{\frac{\pi}{2}}C\ \Gamma(3+n)
\frac{{\rm sin}{[(2+n)(\pi/2)]}}{(2+n)(\pi/2)} r^{-(3+n)}.
\label{eqn:powerlaw}
\end{equation}
This power law fit to the $P(k)$ of LBG clustering is plotted in 
Figure \ref{lbg}.  The Giavalisco \etal (1998) analysis only uses
galaxy pairs with angular separations less than 330 arcsecs, so
in Figure~\ref{lbg} we plot the inferred galaxy $P(k)$ only out to
$k=2\pi/r_{\rm max}$, where $r_{\rm max}$ is the comoving lengthscale
corresponding to 330 arcsecs for the assumed cosmology.

Using a counts-in-cells analysis of a sample with full redshift 
information, Adelberger \etal (1998) estimate a higher amplitude
of LBG clustering, corresponding to $r_0=4 \pm 1\hmpc$ for $\Omega_0=1$
and $\gamma=-1.8$.  This analysis uses cells of quite large comoving
volume ($8\hmpc$ cubes for $\Omega_0=1$), which would be influenced by
fluctuations on scales larger than those probed by our $P(k)$ measurement.
We therefore plot only the Giavalisco \etal (1998) results in
Figure~\ref{lbg}, with the proviso 
that they are likely to be slightly low in amplitude.

The mass $P(k)$ in Figure \ref{lbg} is from 
the fiducial sample, with a mean redshift $z=2.5$.
We rescale $P(k)$ assuming linear growth to the redshift $z=3.04$ of the LBG
results, for two different cosmological models.
The redshift extrapolation is slightly different for the two cosmologies,
but the primary influence of cosmological parameters is on
the conversion from $\kms$ to the comoving $\hmpc$ units used in
Figure~\ref{lbg}.  The parameters have a similar but not identical
influence on the conversion of angular separations to comoving $\hmpc$,
so although the mass and galaxy power spectra are both different in 
the two panels of Figure~\ref{lbg}, the offset between them is nearly the 
same.  

The mass distribution is significantly non-linear on these
scales even at $z=3$, but the $P(k)$ measured from the \lya\ forest is 
representative of the primordial, linear $P(k)$ for the reasons
given in Section 2.2.  From the plots, 
it is evident that the primordial $P(k)$ is steeper than 
the LBG $P(k)$ (with a logarithmic slope of $-2.25$ rather than $-1.2$). 
The difference in slope is caused at least partly by non-linear 
evolution, during which a transfer of power from large to small scales 
tends to make $P(k)$ of this type shallower 
(see, e.g., Baugh \& Efstathiou 1994).  The dashed lines show
the non-linear mass $P(k)$ computed from
the three-dimensional power spectrum of the mass distribution in the
normalizing simulations. We interpolate between the two outputs closest to
$z=3.04$, assuming that the correct amplitude has been set by the \lya\ results
at $z=2.5$. 
The non-linear mass $P(k)$ is indeed shallower than the primordial one,
but it is still at least marginally
steeper than the LBG $P(k)$. 

Despite the statistical uncertainties in both sets of measurements,
it is clear that the galaxy clustering is substantially biased
with respect to the mass clustering.  The ratio of the power spectra
is $\sim 3-10$, depending on scale, which translates to a bias factor
$b \sim 2-3$.  If we adopt the larger LBG clustering amplitude found 
by Adelberger \etal (1998), then the implied bias factors are about
a factor of two larger.  The scale dependence of the bias implied by
Figure~\ref{lbg} is not unreasonable since the scales probed are in
the non-linear regime.  However, it is also possible that we have
overestimated the steepness of the non-linear mass $P(k)$ because
of the finite volume of our normalizing simulations, or that we have
incorrectly estimated the slope of the LBG $P(k)$ because the inversion
equation~(\ref{eqn:powerlaw}) assumes that $\xi(r)$ is a power
law on all scales that contribute significantly to the measured $P(k)$.

Measurements of the bias factor such as this one should be useful in
constraining theories of galaxy formation, especially as 
high-$z$ galaxy samples and \lya\ forest samples
increase in size and the statistical uncertainties
become smaller. One can ask how our direct estimate of $b$ compares with 
those made by Giavalisco \etal (1998) and Adelberger \etal (1998),
who set the amplitude of mass fluctuations by requiring that the
correct masses of clusters be reproduced at $z=0$ (see, e.g., 
White \etal 1993). 
Because the $z=0$ normalization depends on $\Omega_0$ and the
extrapolation from $z=0$ to $z=3.04$ depends on $\Omega_0$ and $\Lambda_0$,
the value of $b$ inferred in this way depends strongly on the adopted
cosmological parameters.
For an open $\Omega_0=0.2$ model, the LBG
results require a relatively low $b \sim 2$ (\cite{adelberger98}).
For $\Omega=1$, a bias factor of 6 or even higher is required.
Since the value of $b$ inferred from Figure~\ref{lbg} 
lies in between these two extremes,  
it seems that our measurement favors an intermediate value of $\Omega_0$.
This constraint can be derived more cleanly by comparing our mass $P(k)$
at $z=2.5$ directly to the combination of $\Omega_0$ and the $P(k)$
amplitude constrained by clusters at $z=0$, as discussed by
Weinberg et al.\ (1998a).

Theoretical models of the LBG population consistently predict strong
bias between LBGs and mass, whether they are based on analytic
approximations (e.g., \cite{mof96}; \cite{adelberger98}; \cite{baugh98}; 
\cite{coles98}), the clustering of massive halos in N-body simulations
(e.g., \cite{bagla98a}b; \cite{colin98};
\cite{jing98}; \cite{wechsler98}),
the combination of N-body simulations with semi-analytic galaxy
formation models (\cite{governato98}; \cite{kauffmann98b}), 
or full hydrodynamic simulations of the LBG population (\cite{khw98}).
In detail, the predictions of the bias and its dependence on
galaxy luminosity depend on the way that LBGs populate their
parent dark halos, on the relation between an LBG's star
formation rate and its mass, and on other aspects of the theory
of galaxy formation.  While many models are consistent with current
estimates of the LBG bias (including the estimate presented here),
more precise measurements of the bias from future \lya\ forest
and LBG data should help to constrain the mechanisms of galaxy 
formation and the nature of LBGs.

\begin{figure*}[t]
\centering
\vspace{7.8cm}
\includegraphics{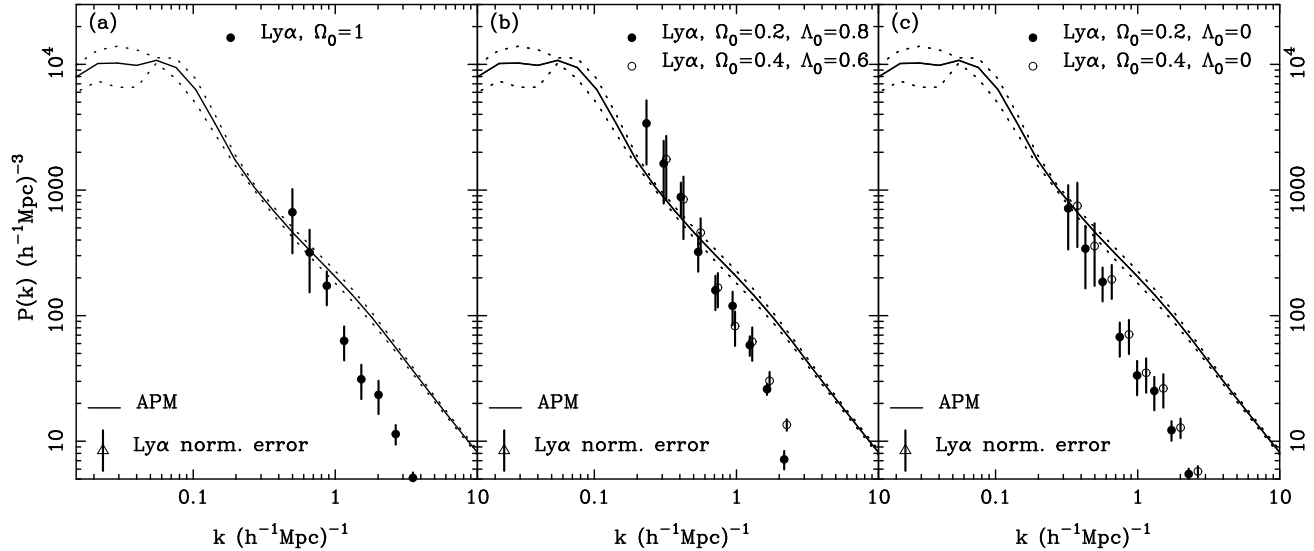}
\caption[junk]{
The normalized $P(k)$ compared to the APM $P(k)$ inverted from 
angular clustering
(\cite{be93}) at $z=0$. We show results obtained after using five different 
cosmologies to do the extrapolation from $z=2.5$ to $z=0$.
Note that these plots compare a linear mass power
spectrum to a non-linear galaxy power spectrum. 
The differences between them illustrate
the advantages of using our measurement of the primordial $P(k)$ 
to test theories directly, despite its large statistical uncertainty.
If any one of the background cosmologies
in these three panels is the correct one, then it is likely 
that both non-linear matter evolution
and galaxy formation physics must be invoked in 
order to explain the galaxy power spectrum results.
\label{apm}}
\end{figure*}

If we want to compare our estimate of $P(k)$ to the power spectrum of low-$z$
galaxy samples, then our choice of background cosmology makes a 
significant difference. 
For different values of $\Omega_{0}$ and $\Lambda_{0}$,
$\kms$ units at $z=2.5$ map to very different length scales at $z=0$ (following
equation [10]). For example, the largest scale on which we can measure 
what we believe is the true $P(k)$ is $k=2.7 \times 10^{-3} \invkms$,
which corresponds to a wavelength of $12 \hmpc$ if $\Omega_{0}=1$ but to
$35 \hmpc$ for an accelerating universe with 
$\Omega_{0}=0.2$ and $\Lambda_{0}=0.8$.
The values of $\Omega_0$ and $\Lambda_0$ also affect the linear
growth factor over this redshift range.

In Figure \ref{apm} we plot the power spectrum of APM galaxies,
recovered from an inversion of their angular clustering by Baugh \& Efstathiou
(1993, the data values themselves are taken from the table in 
Gazta\~{n}aga \& Baugh 1998). We also plot our linear $P(k)$ results
from the \lya\ forest for three different sets of background cosmologies,
an $\Omega_{0}=1$ model (\ref{apm}a), two flat, non-zero $\Lambda$ models
(\ref{apm}b) and two open models (\ref{apm}c). If we compare the shapes and
amplitudes of the
power spectra, we can see that the non-zero $\Lambda$ cosmologies appear to 
prefer some antibias around $k \sim0.2-0.4 \invhmpc$. The error bars are   
large enough that this evidence
is merely suggestive at present. The dip below a power
law seen in the APM $P(k)$ on these scales has not been clearly seen in 
measurements of the power spectrum from galaxy redshift surveys (see, e.g.,
Vogeley 1998). If measurements from new, larger galaxy  surveys confirm
that  it is a real feature, then we could try to look for it in 
future \lya\ $P(k)$ measurements, but it would probably
only be at an accessible range of
scales in the case of a non-zero $\Lambda$ Universe.
We should bear in mind that Figure~\ref{apm} again compares the linear mass
$P(k)$ to the non-linear galaxy $P(k)$. 
Mode coupling is likely to have made the
non-linear $P(k)$ significantly shallower on
scales up to $k \sim 0.1 \invhmpc$
(Baugh \& Efstathiou 1994; Croft \& Gazta\~{n}aga 1998).
The cosmologies used in Figure \ref{apm}a and \ref{apm}c
may therefore be consistent with little or no galaxy bias on 
the scales of the \lya\ forest measurement, and non-linear evolution
might reconcile the shapes of the power spectra for
the non-zero $\Lambda$ cosmologies.

\section{Summary and discussion}
We have presented an estimate of the primordial power spectrum 
of mass fluctuations, $P(k)$. To arrive at this estimate, we have applied the
reconstruction method of CWKH to a set of 19 QSO spectra,
originally measured for other purposes.
The method assumes that the primordial density fluctuations are Gaussian,
as predicted by inflation, and that \lya\ forest absorption
arises predominantly in the diffuse, photoionized IGM, as predicted
by hydrodynamic cosmological simulations.
This physical picture of the \lya\ forest generically predicts a simple,
non-linear relation between the mass overdensity and the \lya\ optical depth.
The one uncertain parameter in this relation (the constant $A$ of
equation~[2]) can be fixed observationally by matching the
mean opacity of the forest ($\taueff$), yielding a $P(k)$ measurement
with no unknown ``bias factors.''

Our measurement of $P(k)$ is made at a mean redshift $z=2.5$
and spans scales from $k= 1.4 \times 10^{-2} - 
2.7\times 10^{-3} \invkms$, which correspond
to  wavelengths of $2 - 12$ comoving$\hmpc$ if $\Omega_0=1$.
Fitting a power law to the data points, we find
a logarithmic slope $n=-2.25 \pm 0.18$.
The amplitude expressed in terms of the variance in the density 
field per unit interval in $\ln k$
is $\Delta^{2}_{\rho}(k_p)\equiv k_p^3 P(k_p)/2\pi^2=0.57^{+0.23}_{-0.17}$.
Here $P_{p}$ is the value of $P(k)$ at
a pivot wavenumber $k_{p}=8 \times 10^{-3} \invkms$,
chosen so that the statistical errors in $n$ and $P(k_p)$ are uncorrelated.
The uncertainty in the amplitude of $P(k_p)$ corresponds to a $1\sigma$
uncertainty of about 18\% in the rms amplitude of mass fluctuations
on the scale $2\pi/k_p \sim 700\kms$.
This error estimate includes the statistical uncertainty in $\taueff$
quoted by PRS; the impact of alternative determinations of $\taueff$
can be found from Figure~\ref{taubar}.

There are a number of reasons for thinking that this measurement
of $P(k)$ is robust, in the sense that any systematic errors are
no larger than our $1\sigma$ statistical errors.
First, the tests of CWKH show that our method successfully
recovers the true linear $P(k)$ from full hydrodynamic simulations
of three different cosmological models, even using artificial
spectra of moderate resolution and signal-to-noise ratio.
Second, we have examined (in Figures~\ref{contfitpk} and~\ref{vary})
the effects of changing the operational parameters used in our
data preparation procedure.  We find that continuum fitting 
uncertainties set an upper limit to the scale on which we can
measure $P(k)$, at $k \sim 2\times 10^{-3}\invkms$.  On smaller
scales, reasonable variations on our standard procedure do not
influence our results at the $1\sigma$ level.
Third, we have examined (in Section 4) the most obvious potential
source of ``spurious'' fluctuations in the \lya\ forest,
spatial variations in the UV background intensity, and shown that
they should have negligible impact on $P(k)$ on the scales accessible
with our current data.  Fourth, the $P(k)$ determined separately
from the low redshift and high redshift subsamples of the data are
consistent with the hypothesis of a single underlying power spectrum
experiencing linear growth from $z=2.75$ to $z=2.1$, albeit with
large statistical uncertainties owing to the smaller size of 
the subsamples (Figure~\ref{zevol}).  Fifth, the $P(k)$
determined from the low-$z$ subsample is consistent with the
$P(k)$ determined from an entirely independent set of nine
QSO spectra (the secondary sample described in Section 2.1)
with the same mean redshift (Figure~\ref{zevol}).
Our $P(k)$ measurement also agrees well with the measurement
presented in CWKH from Songaila \& Cowie's (1996)
Keck HIRES spectrum of Q1422+231, which has
a mean absorption redshift $z=3.2$.  The statistical
precision of the present measurement is much higher than that
of the Q1422+231 measurement because of the greater number
of QSOs that contribute to it.

There are two general ways that our $P(k)$ measurement can be
checked using existing or readily obtainable \lya\ forest data.
The first is simply to confirm the result with independent data sets,
preferably ones that have larger numbers of QSO spectra and hence
yield smaller statistical error bars.  With a larger data set,
one could also carry out a more exacting version of the redshift
evolution test illustrated in Figure~\ref{zevol}.  Our present
data are consistent with linear growth of $P(k)$, but positive detection
of the expected growth (not possible with our current statistical
errors) would be a strong empirical indication that the $P(k)$
derived from the \lya\ forest indeed represents fluctuations in
an evolving mass density field.
The second general approach is to test the predictions of a model
with Gaussian initial conditions and our derived $P(k)$ against
other statistical properties of the \lya\ forest, using high resolution
spectra.  For example, the flux decrement distribution function
(\cite{rauch97}) depends mainly on the amplitude of $P(k)$ and on the
PDF (Gaussian vs. non-Gaussian) of
the primordial fluctuations (Weinberg et al., in preparation).
This statistic can therefore be used to test our $P(k)$ determination
and to test the theoretical assumption that is critical to our
method, the hypothesis of Gaussian initial conditions.
At a greater level of detail, one can check that spatial variations
in statistical properties of the forest (from QSO to QSO or within
individual spectra) are consistent with expectations, 
to constrain any coherent spatial
fluctuations in the temperature-density relation.
Finally, larger samples of high resolution spectra can be used to
measure $\taueff$ (as in \cite{rauch97}), better constraining the
observational parameter used in our $P(k)$ normalization.

Comparison of our derived $P(k)$ to the measured clustering of Lyman
Break Galaxies implies that the latter are a highly biased population,
with a bias factor $b \sim 2-5$.  While the statistical errors in 
both the \lya\ $P(k)$ and the LBG $P(k)$ are presently large, this is
arguably the most direct measurement of bias between galaxies and
mass to date.  The bias factors inferred from comparisons of
galaxy density and peculiar velocity fields (\cite{strauss95} and
references therein) or from redshift-space distortion analyses
(\cite{hamilton98} and references therein) depend strongly on
the assumed value of $\Omega_0$, roughly $b \propto \Omega_0^{0.6}$.
The bias measurement presented here is only weakly dependent
on cosmological parameters, as one can see by comparing 
Figures~\ref{lbg}a and~\ref{lbg}b.  The comparison between our
derived $P(k)$ and the power spectrum of present day galaxies
does depend on cosmology (Figure~\ref{apm}), since the values of
$\Omega_0$ and $\Lambda_0$ affect the amount of fluctuation
growth and the change in velocity scales over the large redshift
interval from $z=2.5$ to $z=0$.

The slope and amplitude of the derived $P(k)$ are consistent with
the predictions of some scale-invariant, COBE-normalized CDM models
(e.g., the OCDM model in Figure~\ref{models}, with $\Omega_0=0.4$,
$h=0.65$, $\sigma_8=0.75$) and inconsistent with others
(e.g., the CCDM model, with $\Omega_0=1$, $h=0.5$, $\sigma_8=1.2$).
As we show in a separate paper (Phillips \etal 1998), COBE-normalized
CDM models with a variety of $\Omega_0$ and $h$ values can fit the \lya\ 
$P(k)$ if the primordial spectral index $n_p$ is treated as a free
parameter, but within any given class of models (e.g., open CDM)
one obtains a constraint on a parameter combination of the form
$\Omega_0 h^\alpha n_p^\beta \Omega_b^\gamma$.  In Weinberg \etal (1998a)
we show that consistency between our $P(k)$ derived at $z=2.5$ and constraints
from the cluster mass function at $z=0$ require a low value of $\Omega_0$
($\Omega_0 \approx 0.45$ for $\Lambda_0=0$ and $\Omega_0 \approx 0.35$
for $\Lambda_0=1-\Omega_0$) if the power spectrum has the large scale
shape implied by studies of galaxy clustering.
Perhaps the most significant theoretical 
implication of our results, already evident in Figure~\ref{models},
is that inflation $+$ CDM models, originally motivated
by considerations of microwave background anisotropies at
$z \sim 1000$ and large scale structure at $z \sim 0$, predict a
$P(k)$ that is at least roughly consistent with our measurement, even
though it probes a different epoch of cosmic history 
and is based on a complex analysis of entirely different observational
phenomena.

The main requirement for improving the precision of our $P(k)$ measurement
is the analysis of a larger sample of QSO spectra.
With 100 full \lya\ forest spectra, it should be possible to reduce
the statistical uncertainty in the amplitude of $P(k)$ below 10\%,
provided that $\taueff$ is determined with sufficient precision 
from high resolution spectra.  Greater statistical precision will merit a
more detailed examination of some potential systematic errors, and it
will also be worth testing continuum fitting procedures on large simulations
to see if the $P(k)$ determination can be extended to larger scales.
In the slightly more distant future, analysis of spectra towards pairs
or close multiples of QSOs can be used to measure redshift-space
distortions of clustering and thereby constrain spacetime geometry,
as proposed by CWKH, Hui, Stebbins, \& Burles (1998), and
McDonald \& Miralda-Escud\'e (1998).  Eventually, the giant samples
of QSO spectra from the 2dF and Sloan redshift surveys may yield
a truly three-dimensional view of evolving large scale structure in the 
intergalactic medium.
The moderate spectral resolution of these samples ($\sim 8$\AA\ and
$\sim 2.5$\AA, respectively) is not in itself an obstacle to such 
a program: as our results here show, by treating each spectrum as a
continuous map instead of a collection of lines, one can measure
large scale fluctuations without resolving small scale features.
The important question will be whether the unabsorbed continuum
can be determined with sufficient accuracy 
from such data over scales larger than the
typical transverse separations of QSO lines of sight.

The power of the \lya\ forest as a test of cosmological theories
derives from the simplicity of the physics that governs the absorbing
medium and from the existence of an observable quantity, $\taueff$,
that calibrates the relation between underlying mass fluctuations
and the observable fluctuations in QSO flux.  The results presented
in this paper illustrate the promise of studies that use the \lya\ 
forest to trace the formation of structure in the Universe.

\bigskip
\acknowledgments

We thank Andrew Gould for helpful discussions and Vijay Narayanan
for assistance with some PM calculations.
This work was supported by NASA Astrophysical Theory Grants
NAG5-3111, NAG5-3922, and NAG5-3820,
by NASA Long-Term Space Astrophysics Grant NAG5-3525,
and by the NSF under grants AST98-02568 and ASC93-18185.


\end{document}